\documentclass[10pt,twocolumn,letterpaper]{article}

\usepackage[pagenumbers]{iccv} % To force page numbers, e.g. for an arXiv version

\usepackage{booktabs}
\usepackage{multirow}
\usepackage{makecell}
\usepackage{svg}
\usepackage{float}

\definecolor{iccvblue}{rgb}{0.21,0.49,0.74}
\usepackage[pagebackref,breaklinks,colorlinks,allcolors=iccvblue]{hyperref}

\def\paperID{20579} % *** Enter the Paper ID here
\def\confName{CVPR}
\def\confYear{2026}

\title{Hybrid Rendering for Multimodal Autonomous Driving: Merging Neural and Physics-Based Simulation}

\author{Máté Tóth\\
aiMotive\\
\and
Péter Kovács\\
aiMotive\\
\and
Réka Bencses \\
aiMotive\\
\and
Zoltán Bendefy\\
aiMotive\\
\and
Zoltán Hortsin\\
aiMotive\\
\and
Balázs Teréki\\
aiMotive\\
\and
Tamás Matuszka\\ 
aiMotive\\
}

\begin{document}

\maketitle

\begin{abstract}
Neural reconstruction models for autonomous driving simulation have made significant strides in recent years, with dynamic models becoming increasingly prevalent. However, these models are typically limited to handling in-domain objects closely following their original trajectories. We introduce a hybrid approach that combines the strengths of neural reconstruction with physics-based rendering. This method enables the virtual placement of traditional mesh-based dynamic agents at arbitrary locations, adjustments to environmental conditions, and rendering from novel camera viewpoints. Our approach significantly enhances novel view synthesis quality—especially for road surfaces and lane markings—while maintaining interactive frame rates through our novel training method, NeRF2GS. This technique leverages the superior generalization capabilities of NeRF-based methods and the real-time rendering speed of 3D Gaussian Splatting (3DGS). We achieve this by training a customized NeRF model on the original images with depth regularization derived from a noisy LiDAR point cloud, then using it as a teacher model for 3DGS training. This process ensures accurate depth, surface normals, and camera appearance modeling as supervision. With our block-based training parallelization, the method can handle large-scale reconstructions ($\geq 100 000$ $m^2$) and predict segmentation masks, surface normals, and depth maps. During simulation, it supports a rasterization-based rendering backend with depth-based composition and multiple camera models for real-time camera simulation, as well as a ray-traced backend for precise LiDAR simulation.

\end{abstract}    
\section{Introduction}
The development of robust autonomous driving systems depends heavily on diverse datasets for training and evaluation. Traditional approaches rely on recordings of real-world driving scenarios. However, datasets such as \cite{caesar2020nuscenes, sun2020scalability, chang2019argoverse, matuszka2023aimotive} often lack critical safety-related edge cases due to their rarity. Capturing these infrequent events is both costly and logistically challenging, making alternative approaches necessary. As a result, there is growing interest in synthetic data generation techniques, which offer a promising solution to bridge the gap in safety-critical scenarios for autonomous vehicle development.

Traditional simulation methods for autonomous driving, such as CARLA \cite{dosovitskiy2017carla}, are based on principles similar to those used in 3D game engines. They rely on mesh-based 3D models and traditional computer graphics rendering techniques, such as rasterization or ray tracing. While these simulators provide a high degree of control over the simulated environment, they have two major drawbacks. First, creating detailed 3D environments and assets requires extensive manual effort. Second, an unavoidable domain gap remains between simulated and real-world data.

Neural Radiance Fields (NeRF) \cite{mildenhall2020nerf} significantly improved reconstruction quality compared to earlier data-driven 3D scene reconstruction methods. More recently, 3D Gaussian Splatting (3DGS) \cite{kerbl3Dgaussians} introduced an explicit scene representation that enables real-time rendering while maintaining near-photorealistic reconstruction quality. These advancements have made data-driven neural reconstruction methods a viable option for automotive simulation.

Several adaptations have been made to 3DGS for automotive applications, primarily focusing on reconstructing dynamic scenes using supervised and unsupervised settings and integrating additional sensor modalities used in autonomous driving, such as LiDAR. However, these methods still lack one of the key advantages of traditional simulation techniques: controllability. Inserting unseen objects or repositioning objects far from their original trajectory remains challenging, often leading to degraded reconstruction quality and limiting their effectiveness for autonomous driving downstream tasks.

In this paper, we present a hybrid rendering method that combines the strengths of neural reconstruction and traditional computer graphics to overcome their respective limitations. Our approach reconstructs the static environment using neural rendering while incorporating dynamic actors through conventional graphics techniques. This enables flexible placement of arbitrary objects, adjustable weather conditions, and rendering from novel viewpoints. By leveraging neural rendering, we reduce the domain gap commonly associated with traditional simulators while preserving their controllability. This allows us to mitigate the weaknesses of both methods. Additionally, our solution supports multimodal output, including camera sensor simulation, LiDAR point clouds, depth maps, and surface normals. To demonstrate the effectiveness of our approach, we conduct experiments on data generated by our hybrid method, evaluating its suitability for autonomous driving development. In summary, our main contributions are the following:
\begin{itemize}
    \item A hybrid rendering approach that combines neural reconstruction and traditional computer graphics techniques to overcome their limitations.
    \item Enhanced novel view synthesis introducing NeRF2GS method.
%    \item A novel block generation algorithm for large-scale reconstruction.
    \item Benchmark results for downstream autonomous driving tasks using data generated by our method.
\end{itemize}
\section{Related Work}
\label{sec:rel_work}

%Several efforts have been made to adapt neural reconstruction methods to the needs of autotmotive simulation. One of the most important challenges in scene reconstruction is the handling of dynamic objects in the scene. Current works approach this by trying to reconstruct the original dynamic object of the recording using supervised or unsupervised learning techniques. NeRF-based supervised methods, such as Neural Scene Graphs \cite{ost2021neural}, MARS \cite{wu2023mars} and NeuRAD \cite{tonderski2024neurad} use bounding box data to to learn reconstruct the dynamic objects as static ones in a canonical space, and transform them back to their world-space locations. Meanwhile the unsupervised EmerNeRF \cite{yang2023emernerf} method aims to reconstruct the dynamic objects using a temporally deformable dynamic field. While some of these methods offer convvincing reconstruction quality, they are not suitable for interactive simulator applications due to the slow rendering speeds of NeRFs.

Many efforts have been made to adapt neural reconstruction methods for automotive simulation. One of the key challenges in scene reconstruction is dealing with dynamic objects. Current approaches address this issue by attempting to reconstruct the original dynamic objects in the scene using either supervised or unsupervised learning techniques. Supervised NeRF-based methods, such as Neural Scene Graphs \cite{ost2021neural}, MARS \cite{wu2023mars}, and NeuRAD \cite{tonderski2024neurad}, use bounding box data to learn how to reconstruct dynamic objects as static ones in a canonical space, and then transform them back to their original positions in world space. On the other hand, the unsupervised EmerNeRF \cite{yang2023emernerf} method focuses on reconstructing dynamic objects using a temporally deformable dynamic field. While some of these methods achieve impressive reconstruction quality, they are not suitable for interactive simulator applications due to the slow rendering speeds of NeRFs.

%This limitation is largely solved by 3DGS-based reconstruction methods such as the unsupervised S3Gaussian \cite{huang2024textit}, Periodic Vibration Gaussian \cite{chen2023periodic} or DeSiRe-GS \cite{peng2024desire} achieve excellent reconstruction results for dynamic scenes at the trainig trajectories, however their controllability (in terms of dynamic actor manipulation) is severely limited, and they only support RGB camera simulation which is not always adequate for AD aplications.
%Supervised 3DGS methods, like Street Gaussians \cite{yan2024street}, DrivingGaussian \cite{Zhou_2024_CVPR} and OmniRe \cite{chen2024omnire}, offer some manipulation capabilty for dynamic objects, but it is limited to slight changes to their positions and rotations, and they also do not support LiDAR simulation or arbitrary camera models.
%SplatAD \cite{hess2024splatad} solves Lidar simulation within the 3DGS framework, but the other mentioned limitations still persist.

This limitation is largely addressed by 3DGS-based reconstruction methods such as the unsupervised S3Gaussian \cite{huang2024textit}, Periodic Vibration Gaussian \cite{chen2023periodic}, and DeSiRe-GS \cite{peng2024desire}, which achieve high-quality reconstructions of dynamic scenes along the training trajectories. However, their ability to manipulate dynamic actors is severely restricted, and they only support RGB camera simulation, which is often insufficient for autonomous driving applications. 

Supervised 3DGS methods, including Street Gaussians \cite{yan2024street}, DrivingGaussian \cite{Zhou_2024_CVPR}, and OmniRe \cite{chen2024omnire}, provide some control over dynamic objects, allowing minor adjustments to their positions and rotations, but it is still limited. Additionally, they do not support LiDAR simulation or arbitrary camera models. SplatAD \cite{hess2024splatad} extends 3DGS-based methods by incorporating LiDAR simulation, but the other aforementioned limitations remain unresolved.

%\TOD{Related work is gonna be here.
%Include Block-NeRF and highlight why their block generation algorithm is insufficient for large-scale reconstructions with complex ego-trajectory.}
\begin{figure}[t]
  \centering
   \includegraphics[width=1.0\linewidth]{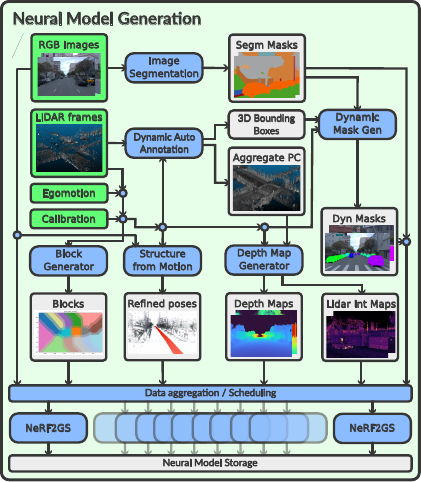}
   \caption{Overview of the reconstruction model generation method.}
   \label{fig:train_pipeline}
\end{figure}

\section{Hybrid Rendering for Autonomous Driving}
\begin{figure}[t]
  \centering
   \includegraphics[width=1.0\linewidth]{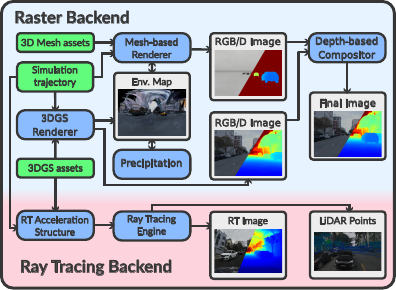}
   \caption{Overview of the rendering pipeline used in our method.}
   \label{fig:render_pipeline}
\end{figure}

Our proposed hybrid rendering method addresses the challenges mentioned in Section \ref{sec:rel_work} by integrating neural reconstruction and a traditional computer graphics rendering pipeline, each composed of multiple specialized submodules. Figure \ref{fig:train_pipeline} depicts the neural reconstruction model generation process, detailing the transformation from raw input to the final model while the high-level overview of the rendering pipeline is visualized in Figure \ref{fig:render_pipeline}.

%Our proposed hybrid rendering method overcoming the above-mentioned issues comprises a neural model generation process and a rendering pipeline, each consisting of multiple corresponding submodules. 

\subsection{Preparing training data for neural reconstruction}
Our pipeline uses four input data sources for the reconstruction: \textit{images} from the high-resolution onboard cameras of the vehicle, \textit{point clouds} generated by one or more onboard LiDARs, \textit{egomotion} and \textit{extrinsic and intrinsic calibration} of the cameras.

\subsubsection{Dynamic object handling}

Our method focuses on highly controllable reconstruction with strong extrapolation capabilities. Since current state-of-the-art approaches can only reconstruct dynamic scenes to a limited extent, we restrict our reconstruction to the temporally consistent, static parts of the scene. To achieve this, we mask out dynamic objects, such as pedestrians and moving vehicles, from the ground truth (GT) images.

First, we use OneFormer \cite{jain2023oneformer} to generate panoptic segmentations for the GT images. We then extract instance masks for relevant objects and track them in image space using an optical flow-based frame-to-frame tracking method. To identify potential dynamic objects, we use a 3D object detection model that processes LiDAR and camera inputs to generate 3D bounding boxes and determine their stationarity in world space. Next, we project LiDAR points from the current frame onto the instance masks, labeling them accordingly. By matching the bounding boxes with instance masks using the labels of enclosed LiDAR points, we effectively track segmentation masks while incorporating stationarity information. Finally, we refine and merge the image-space and bounding box-based tracklets, and create the set of dynamic masks by adding the instance masks corresponding to nonstationary tracklets.

\subsubsection{LiDAR intensity and depth map generation}

Neural reconstruction methods can generate near-photorealistic RGB renderings, but they often struggle to reconstruct accurate geometry in cases of sparse training views or weakly textured and blurred regions, such as road surfaces. Since our method aims to simulate LiDAR sensors, depth supervision and LiDAR intensity rendering are essential.

%To achieve this, we generate dense depth and intensity maps from LiDAR data. We first create a static aggregated point cloud by removing LiDAR points that fall within the bounding boxes of dynamic objects in each sweep, then merge the sweeps using per-point egomotion correction. Next, we generate sparse depth and intensity maps by projecting the point cloud onto the camera frames. We refine the depth maps using a classical depth-completion algorithm \cite{ku2018defense} and apply a simple depth-based dilation to the intensity maps.
To achieve this, we generate dense depth and intensity maps from LiDAR data. We first create a static aggregated point cloud by removing LiDAR points that fall within the bounding boxes of dynamic objects in each sweep, Then merge the sweeps using per-point egomotion correction. Then we split the aggregated point cloud into square chunks in BEV space, and employ Poisson surface reconstruction \cite{kazhdan2006poisson} for each chunk to generate a dense, mesh-based representation of the lidar data. Finally, we use Open3D \cite{Zhou2018} to render the dense depth maps and LiDAR intensity images from these meshes using the intrinsic and extrinsic camera parameters of the train camera frames.  

\subsubsection{Camera pose refinement}
%Menawhile high-fidelity 3D scene reconstruction necessitates highly accurate intrinsic calibration and camera poses, the egomotion-derived world-space camera extrinsics in automotive datasets are often quite noisy. We address this by refining camera poses and intrinsic parameters using a modified version of COLMAP \cite{schoenberger2016sfm}. We use the original calibration as initialization and iteratively apply triangulation and bundle adjustment steps until we reach convergence. Since structure from motion methods are translation, rotation and scale invariant, we have to transform the poses back to the original world-space. To achieve this, we apply the Kabsch algorithm \cite{kabsch1976solution} between the original and optimized camera positions.

While high-fidelity 3D scene reconstruction requires highly accurate intrinsic calibration and camera poses, the egomotion-derived world-space camera extrinsic matrices in automotive datasets are often noisy. To address this, we refine camera poses and intrinsic parameters using a modified version of COLMAP \cite{schoenberger2016sfm}. Starting with the original calibration as initialization, we iteratively apply triangulation and bundle adjustment until convergence.

Since structure-from-motion methods are invariant to translation, rotation, and scale, we must transform the optimized poses back to the original world-space. To achieve this, we apply the Kabsch algorithm \cite{kabsch1976solution} between the original and optimized camera positions.

\subsubsection{Block generation}
%Autonomous driving tasks often need large scenes to be reconstructed in a highly scalable manner, however, neural reconstruction methods do not scale well to very large scenes. To tackle this problem, we opted to decompose the scene to multiple overlapping blocks using a dynamic bird's-eye view (BEV) raster-based block representation. 
%First, we transform the camera positions to a 2D BEV space and iteratively apply k-nearest neighbors clustering with linearly increaing cluster count in each iteration until we get a clustering that satisfies our predefined maximum cluster radius and image count conditions. We discretize the BEV space (creating a 2D raster), and assign each cell to the cluster with the most points lying on it. We extend the clusters to their convex hulls, and assign the previously unassigned cells to the nearest cluster. Finally, we dilate the clusters to achieve a given overlap between them.
%This method allows us to have a guaranteed overlap between the blocks, which is important for reducing popping artifacts at the block edges, while it is flexible enough to efficiently cover irregular train trajectories.  

Autonomous driving tasks often require large-scale scene reconstruction in a highly scalable manner. However, neural reconstruction methods do not scale well to very large scenes. 
There are several approaches, like Mega-NeRF \cite{turki2022mega}, SUDS \cite{turki2023suds}, and Block-NeRF \cite{tancik2022blocknerf} that address this issue using a square grid or clustering-based space partitioning approach, by training separate reconstruction model for each of the blocks.
Our solution is motivated by \cite{turki2023suds} and it is also based on clustering. We decompose the scene into multiple overlapping blocks using a dynamic bird’s-eye view (BEV) raster-based block representation. 

First, we transform the camera positions into a 2D BEV space and iteratively apply k-nearest neighbors clustering, gradually increasing the cluster count in each iteration until the predefined maximum cluster radius and image count conditions are met. We then discretize the BEV space into a 2D raster (where each cell can be contained by multiple blocks at the same time), and assign each cell to the cluster with the highest number of points. Next, we expand the clusters to their convex hulls and assign any remaining unassigned cells to the nearest cluster. Finally, we apply dilation to the clusters to ensure a specified level of overlap.

\begin{figure}[t]
  \centering
   \includegraphics[width=1.0\linewidth]{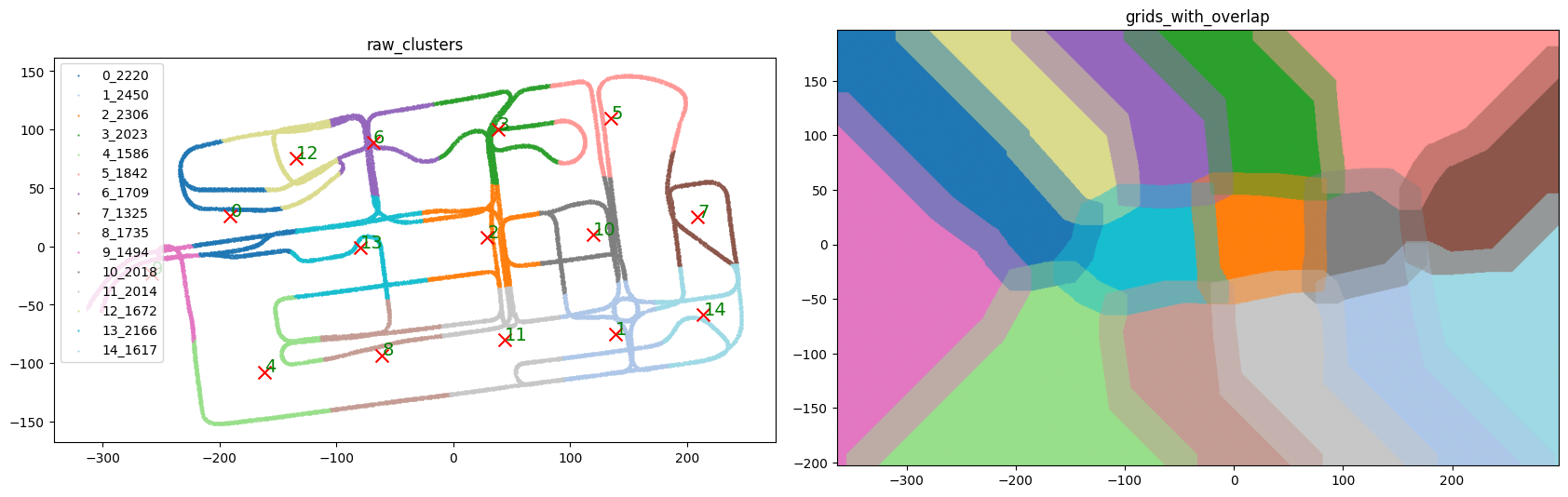}
   \caption{Clustered BEV training poses (left) and the generated blocks with overlaps (right)}
   \label{fig:blocks}
\end{figure}

This approach guarantees overlap between blocks, which is crucial for minimizing popping artifacts at block edges while remaining flexible enough to efficiently cover irregular camera trajectories. It also makes interpolation possible between the distinct reconstruction models based on the cluster edge distances, which greatly reduces artifacts in case of traversing through block boundaries. Figure \ref{fig:blocks} shows the output of the block generation algorithm on a complex scenario.

\subsection{NeRF2GS reconstruction method}

%Explicit, point-based methods such as 3DGS offer comparable reconstruction quality to SOTA implicit methods like Zip-NeRF \cite{barron2023zip} while being able to render at real-time framerates, however they still fall short in terms of extrapolation capability, especially when trained with sparse input views or varying lighting conditions. It has been shown, that training an implicit (NeRF-like) on the original data and using it as initialization and supervision to train a Gaussian-splatting-based model can significantly improve Novel View Synthesis quality \cite{niemeyer2024radsplat}.
%Our reconstruction method (called NeRF2GS) follows this principle: We train a customized on the original RGB images, supervised by the dense LiDAR depth maps and using a learned camera exposure correction method, then we render the outputs of this network at the original training poses and export a point cloud from it to serve as GT data and initialization for training a 3DGS-based reconstruction.

Explicit, point-based methods like 3DGS provide reconstruction quality comparable to state-of-the-art implicit methods such as Zip-NeRF \cite{barron2023zip}, with the added benefit of real-time rendering capabilities. However, they still fall short in terms of extrapolation, particularly when trained with sparse input views or under varying lighting conditions. Research has shown that training an implicit model (such as NeRF) on the original data and using it as both initialization and supervision for training a Gaussian-splatting-based model can significantly enhance Novel View Synthesis quality \cite{niemeyer2024radsplat}. Our reconstruction method, NeRF2GS, builds on this approach: we first train a customized NeRF model on the original RGB images, supervised by dense LiDAR depth maps, and incorporate a learned camera exposure correction method. We then render the outputs of this model at the original training poses and generate a point cloud, which serves as ground truth data and provides initialization for training a 3DGS-based reconstruction.

\subsubsection{NeRF training}
%We use a modified version of the Nerfacto model within the Nerfstudio framework \cite{nerfstudio}. Depth supervision is implemented using the depth loss proposed in \cite{rematas2022urf} with segmentation-based loss weights and an exponentially decaying loss scheduling scheme. We employ an L2 regularization loss on the density accumulation in regions that are classified as the sky in the GT segmentation. This is done to reduce floating artifacts in sky regions with no depth GT. We added a LiDAR intensity prediction network, which is an MLP that takes the geometric feature vector of the Nerfacto density field concatenated with a sinusoidal position encoding and predicts an intensity value. The intensities are supervised using the dense LiDAR intensity maps with an L2 loss. Learned segmentation is also added, however, in this case, we predict a 32-dimensional vector (corresponding to the segmentation classes) that represents the raw logits and use a categorical cross-entropy loss to supervise it on the semantic segmentation maps.

We use a modified version of the Nerfacto model within the Nerfstudio framework \cite{nerfstudio}, incorporating several improvements to enhance reconstruction quality. Depth supervision is implemented using the depth loss proposed in Urban Radiance Fields \cite{rematas2022urf} with segmentation-based loss weighting and an exponentially decaying loss scheduling scheme. To mitigate floating artifacts in sky regions where no depth ground truth is available, we apply L2 regularization to the density accumulation in areas classified as sky in the ground truth segmentation.

Additionally, we introduce a LiDAR intensity prediction network—a multi-layer perceptron (MLP) that takes the geometric feature vector from the Nerfacto density field, concatenated with a sinusoidal position encoding, and predicts an intensity value. This prediction is supervised using dense LiDAR intensity maps with an L2 loss. We also incorporate learned segmentation in a similar manner to \cite{ZhiICCV2021}. We predict a 32-dimensional vector representing raw logits corresponding to segmentation classes, and train it using a categorical cross-entropy loss on the semantic segmentation maps.

%Since automotive recordings usually use automatically adjusted camera ISP parameters (such as auto gain, auto white balance, and local tone mapping), an  "ISP correction" method is needed for sufficient reconstruction quality. To tackle this problem, the most common method is the use of some form us appearance embedding \cite{martin2021nerf}, however we have opted to use a bilateral guided ISP disentanglement method proposed in \cite{wang2024bilateral}.
%At training time we optimize a 3D Bilateral Grid fro each of the training images as \cite{wang2024bilateral} proposed, essentially creating an "ISP enhancement free" scene representation, but for rendering the images for the 3DGS training we do not use any post process step since the ISP post-process step will be handled by the hybrid render engine. 

Automotive recordings typically use cameras with automatically adjusted ISP (Image Signal Processing) parameters, such as auto gain, auto white balance, and local tone mapping, which can introduce inconsistencies in reconstruction. To address this, the common approach is to use some form of appearance embedding \cite{martin2021nerf}. However, we opted for the bilateral guided ISP disentanglement method proposed in \cite{wang2024bilateral}, which better preserves image consistency. During training, we optimize a 3D Bilateral Grid for each training image as described in \cite{wang2024bilateral}, effectively creating an "ISP enhancement-free" scene representation. However, when rendering images for 3DGS training, we omit the post-processing step, as ISP adjustments are handled by the hybrid rendering engine.

%A modified version of the Depth-Nerfacto model with depth loss scheduling, omnidirectional camera model support, and a learned affine color transform-based color correction is used for NeRF training, and a depth-supervised modification of the Splatfacto model is used as the 3DGS method.

\subsubsection{3D Gaussian splatting training}
Our Gaussian Splatting methodology builds upon the established training approach introduced in \cite{kerbl3Dgaussians}, incorporating the enhanced AbsGS densification strategy \cite{ye2024absgs} and utilizing gsplat\cite{ye2024gsplatopensourcelibrarygaussian} as its backend. We employ direct depth and normal regularization to improve model robustness in novel view synthesis and provide these attributes for subsequent stages within our hybrid rendering framework. Additionally, LiDAR intensity and segmentation classes are embedded as additional per-Gaussian features. A preceding NeRF model is leveraged to render color-corrected RGB, depth, normal, LiDAR intensity, and segmentation images, and to augment the initial COLMAP point cloud with a dense point cloud for 3DGS initialization. 

The normal regularization is inspired by the method of DN-Splatter \cite{Turkulainen2024DNSplatterDA}, wherein the normal vector of each Gaussian is defined as the directional vector corresponding to its minimum scale component:
\begin{equation}
    \mathbf{\hat{n}_i} = \mathbf{R_i}\cdot\textrm{OneHot}(\textrm{arg min }(s_{i,1}, s_{i,2}, s_{i,3}))
\end{equation}
where $\mathbf{R_i}$ represents the rotation matrix of the i\textsuperscript{th} Gaussian, and $s_i$ denotes the associated scale vector. These computed normal vectors are then rendered analogously to color channels, and the corresponding loss term is formulated as:
\begin{equation}
    \mathcal{L}_{norm} = -\lambda_{norm}\mathbf{\hat{n}_{pred}\cdot \hat{n}_{gt}}
\end{equation}
To ensure the well-definedness of the normal vectors, we additionally employ a regularization term that encourages the Gaussians to contract along their minimum scale dimension until a predetermined size threshold is reached:
\begin{equation}
    \mathcal{L}_{flat} = \sum_i \textrm{ReLU}(\textrm{min}(s_{i,1}, s_{i,2}, s_{i,3})-C_{flat})
\end{equation}

LiDAR intensity values are incorporated as supplementary features for each Gaussian. These features are normalized to the range [0, 1] via a sigmoid activation function and subsequently rendered as color channels. An L1 loss function is employed to train these features, aiming to reproduce the dense reflectivity images generated by the NeRF model:

\begin{equation}
\label{eq:loss_lidar}
    \mathcal{L}_{lidar} = \lambda_{lidar}|I_{pred}-I_{gt}|
\end{equation}

Segmentation labels are trained analogously to the LiDAR intensities. Our internal 32 segmentation labels are transformed into 6-bit binary representations, and the corresponding six additional Gaussian features are trained independently using the same L1 loss as defined in Equation \ref{eq:loss_lidar}, to replicate these binary values:
\begin{equation}
    \mathcal{L}_{segm} = \sum_{i=0}^6 |S_{pred, i} - S_{gt, i}|
\end{equation}
where $S_i$ represents the image rendered from the i\textsuperscript{th} segmentation feature. During inference, each of the six rendered segmentation bits is rounded to the nearest integer, and the segmentation label ID is reconstructed by interpreting these rounded values as the binary digits of the label identifier.

\subsection{Rendering pipeline}

The original rendering algorithms \cite{zwicker2002ewa, kerbl3Dgaussians} are designed for perspective projection.
Our goal was to generalize the evaluation of 3D Gaussians adapting it to arbitrary sensors; therefore, we needed to avoid any kind of projection.
The result is a small, high performance, easily portable algorithm utilizing the response calculation method based on \cite{moenne20243d}.
Portability lets us use a shared codebase for rasterization and ray tracing for calculating the contribution of a Gaussian along a ray.
This solution also allows us to mitigate popping artifacts and expand the possibilities of neural reconstruction, e.g. with LiDAR simulation.

We integrated this method into our rendering systems, one based on rasterization and one based on ray tracing, where both techniques support arbitrary camera distortion models (fisheye, equirectangular, etc.).

\begin{figure*}
    \centering
    \includegraphics[width=1.0\linewidth]{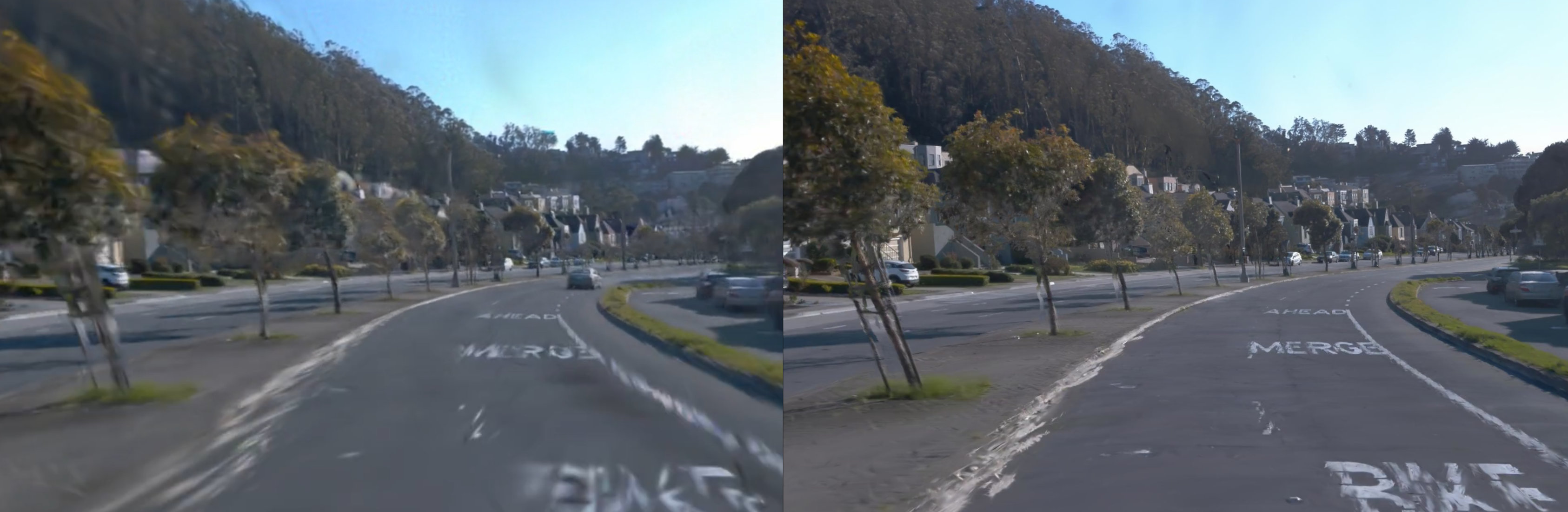}
    \caption{Qualitative comparison of novel view synthesis results between OmniRe (left) and our model (right) for the static components of the scene. Differences in reconstruction quality are particularly noticeable in the rendering of lane boundaries and road markings.}
    \label{fig:omnire_compar_novel_big}
\end{figure*}

\subsubsection{Rasterization}

%Wide-angle or equirectangular camera distortion models require 360-degree information, rendered using six virtual perspective cameras. The images from these virtual cameras must be flawlessly stitched, which is difficult to achieve with the original algorithms, however our solution bypasses this issue.

Wide-angle or equirectangular camera distortion models typically require 360-degree information, which is rendered using six virtual perspective cameras. These camera outputs must be seamlessly stitched, a task that traditional algorithms struggle with. However, our solution bypasses this challenge.

%During rasterization of the Gaussians' 3D proxy geometry we calculate rays for every pixel and evaluate their contribution based on their maximum response along the ray. Since this implementation relies solely on rays intersecting a proxy geometry instead of projecting rectangles, it yields consistent results in any direction and projection, eliminating the stitching artifacts. Consequently, we can prevent popping artifacts by sorting based on the distance from the viewpoint rather than the z-distance from the near planes of the virtual cameras.

In our approach, during the rasterization of the Gaussians' 3D proxy geometry, we calculate rays for each pixel and evaluate their contribution based on the maximum response along the ray. Unlike traditional methods that rely on projecting rectangles, our method focuses solely on rays intersecting the proxy geometry, ensuring consistent results across all directions and projections. This eliminates stitching artifacts, allowing us to avoid popping effects by sorting the data based on distance from the viewpoint rather than the z-distance from the near planes of the virtual cameras.

%Since 3D Gaussians already provide depth information, the rasterization process can continue by rendering dynamic mesh-based objects with traditional depth-testing. For shading the mesh-based objects, we use Image-based lighting (IBL) that captures the lighting information and reflections from the composited environment. If a tessellated ground surface is available, we can render screen space ambient occlusion at the contact area between the mesh and the neural-rendered environment. The depth compositing allows us to add precipitation, like snow or rain to the scene, which is also influenced by IBL.

Since 3D Gaussians already provide depth information, the rasterization process can be continued by rendering dynamic mesh-based objects using traditional depth-testing methods. For shading these mesh objects, we use Image-based Lighting (IBL), which captures the environmental lighting and reflections. When a tessellated ground surface is available, we can render screen space ambient occlusion at the contact area between the mesh and the neural-rendered environment. Additionally, the depth compositing enables us to introduce weather effects, such as snow or rain, which are influenced by the IBL. Finally, the images from the virtual cameras are stitched together into a single image, based on the camera distortion model.

\subsubsection{Ray tracing}

%The ray-tracing algorithm used for rendering the scene is based on \cite{moenne20243d}. This method uses the hardware-accelerated ray-tracing pipeline. Each 3D Gaussian is represented as a proxy geometry in the acceleration structure. Rays are generated in the ray-generation shader according to the camera distortion model or a LiDAR scanning pattern. During the acceleration structure traversal, the splats along a ray are sorted based on distance into a k-sized hit-buffer in the any-hit shaders. The hit-buffer is then processed in the ray-generation shader for computing the individual contribution of the intersected Gaussians based on their maximum response along the ray. In addition to generating reflections, we use surface normal information encoded in the Gaussians to simulate the occlusion of ambient light incoming to the surface.

The ray tracing algorithm employed for scene rendering follows the approach outlined in \cite{moenne20243d}. This method leverages the hardware-accelerated ray tracing pipeline, where each 3D Gaussian is represented as a proxy geometry within the acceleration structure. Rays are generated in the ray-generation shader, based on either the camera distortion model or a LiDAR scanning pattern. During the acceleration structure traversal, the splats along a ray are sorted by distance into a k-sized hit-buffer within the any-hit shaders. This hit-buffer is then processed in the ray-generation shader to compute the individual contributions of the intersected Gaussians, considering their maximum response along the ray. In addition to generating reflections, surface normal information encoded within the Gaussians is used to simulate the occlusion of ambient light incoming to the surface.
\section{Results and Experiments}
%Figure \ref{fig:multimodal} demonstrates that our reconstruction method produces high-quality 3D reconstructions, accurately capturing geometric details and surface properties across RGB, depth, LiDAR intensity, normal, and segmentation modalities.

\begin{figure}[b]
    \centering
    \includegraphics[width=1.0\linewidth]{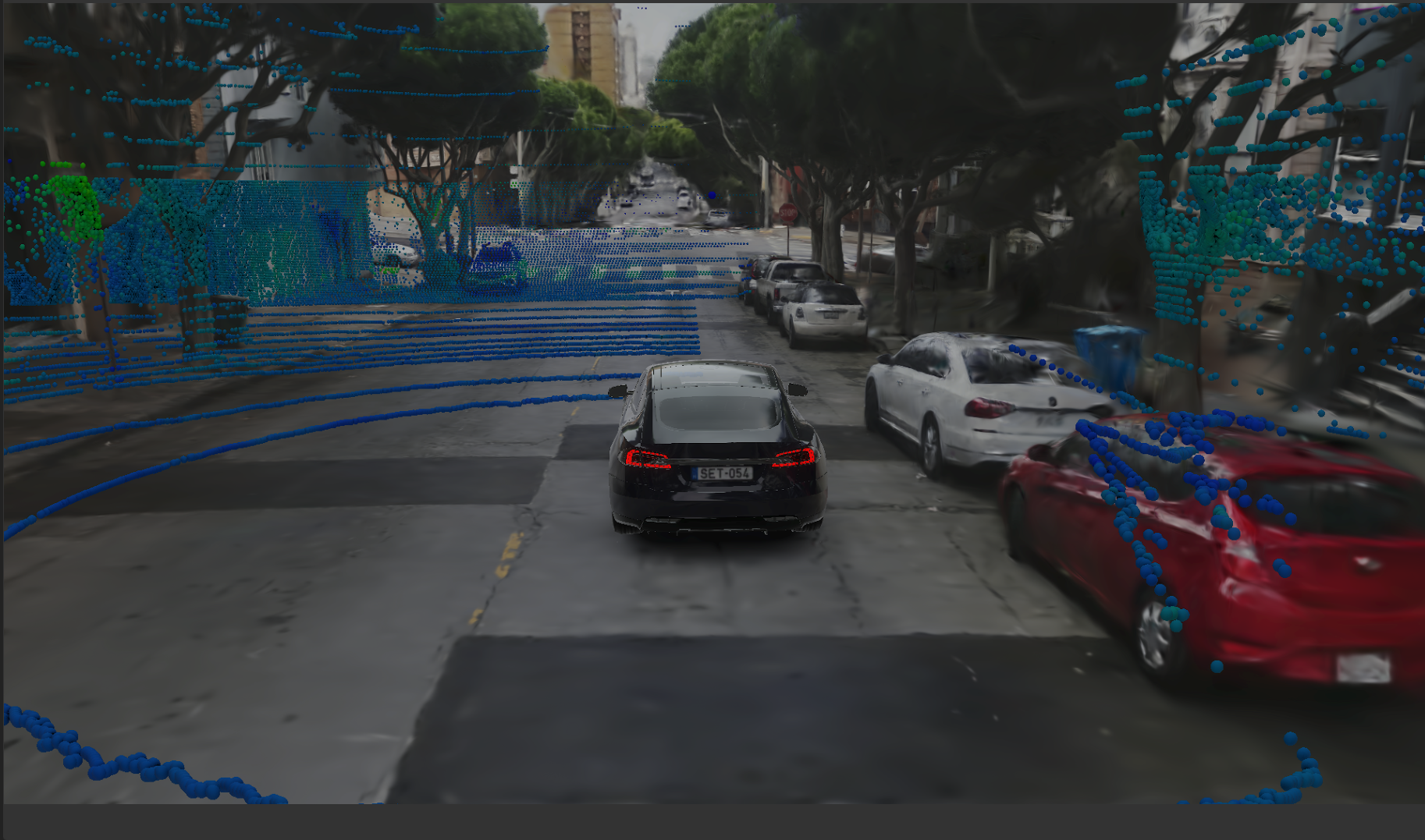}
    \caption{Visualization of the LiDAR point cloud rendered by our method.}
    \label{fig:lidar_fig}
\end{figure}

\begin{figure*}[t]
    \centering
    \begin{subfigure}{0.19\linewidth}
        \centering
        \includegraphics[width=\linewidth]{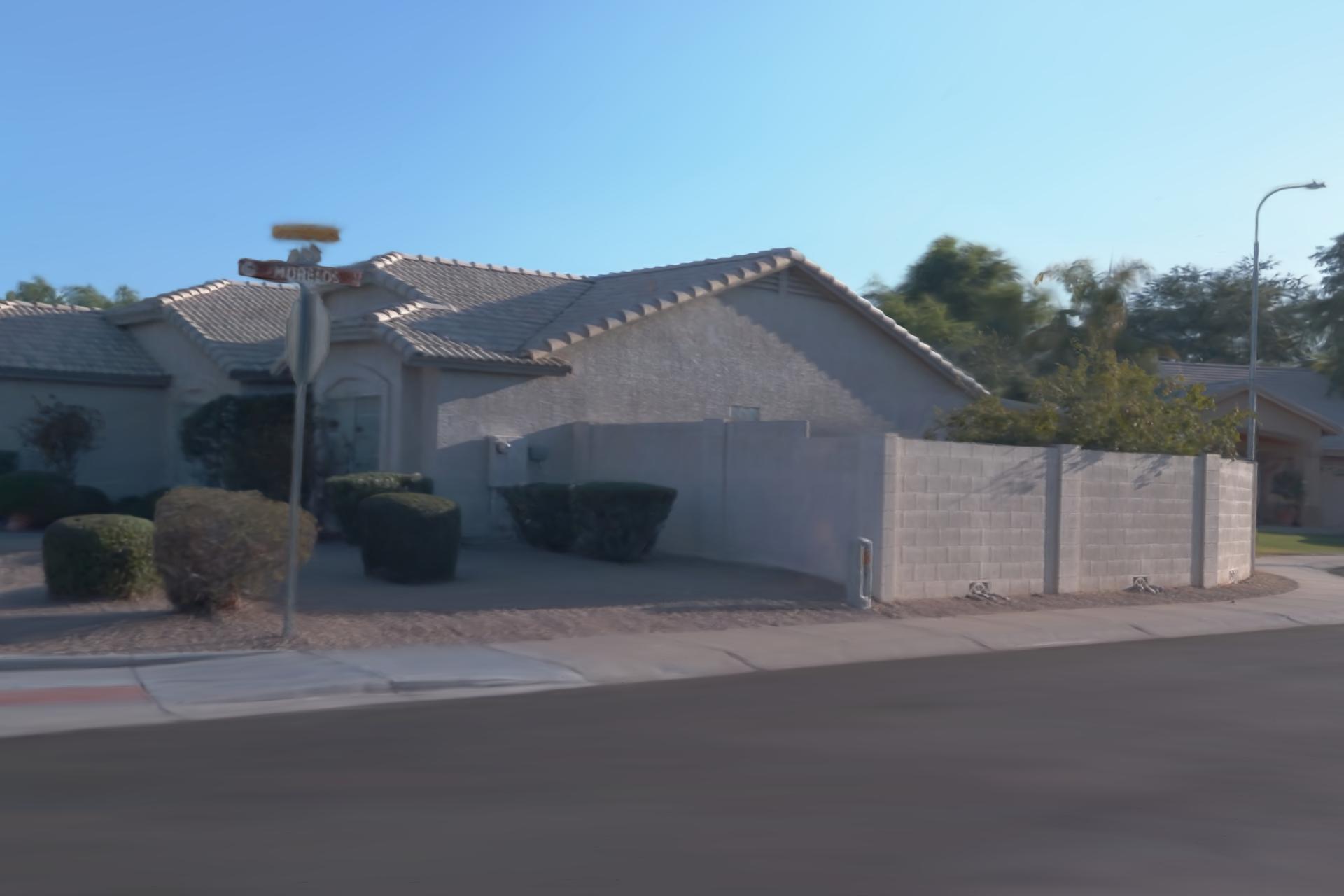}
        \caption[Network2]%
        {{\small RGB reconstruction}}    
        \label{fig:mean and std of net14}
    \end{subfigure}
    \begin{subfigure}{0.19\linewidth}
        \centering
        \includegraphics[width=\linewidth]{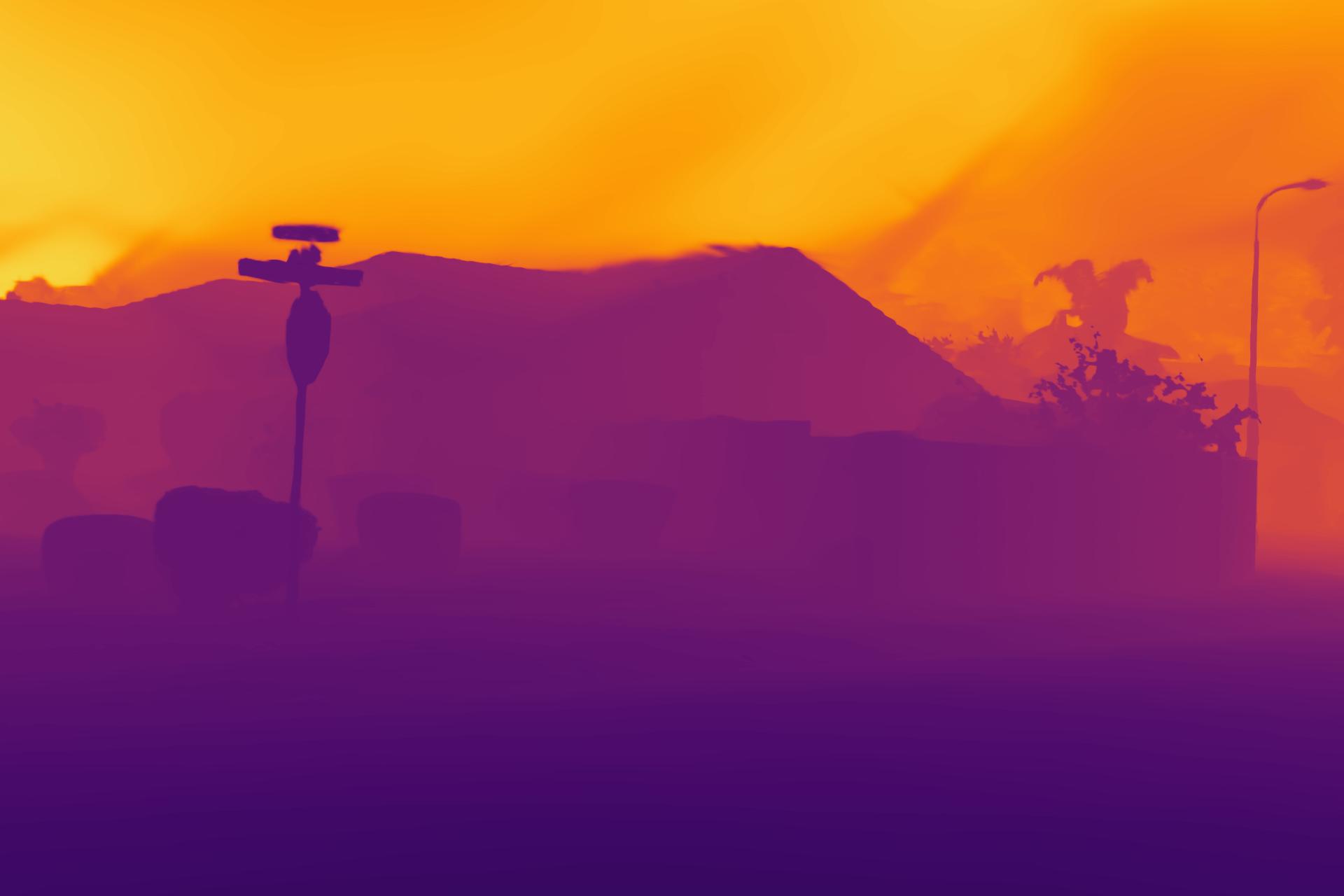}
        \caption[Network2]%
        {{\small Depth image}}    
        \label{fig:mean and std of net14}
    \end{subfigure}
    \begin{subfigure}{0.19\linewidth}
        \centering
        \includegraphics[width=\linewidth]{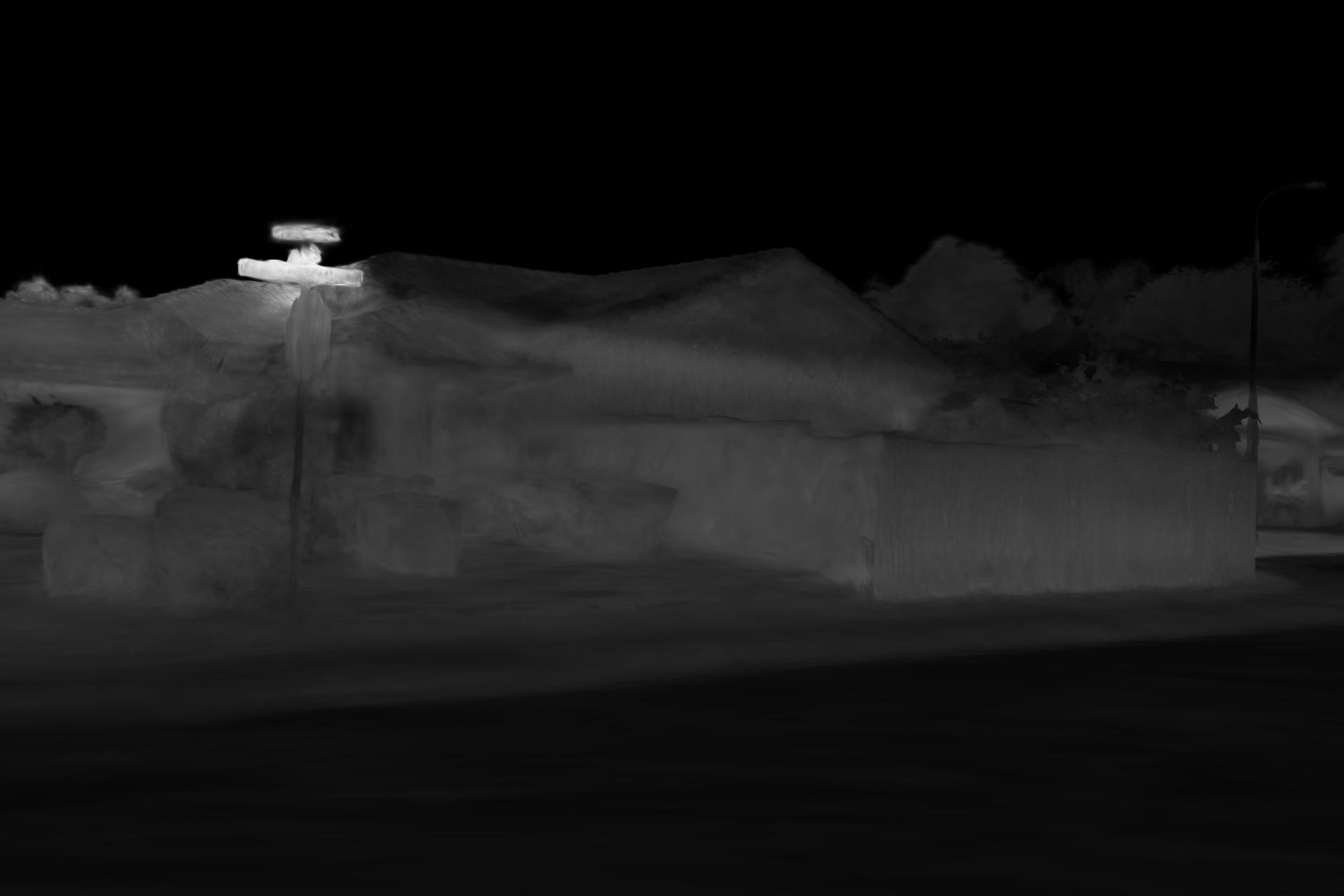}
        \caption[Network2]%
        {{\small Lidar intensities}}    
        \label{fig:mean and std of net14}
    \end{subfigure}
    \begin{subfigure}{0.19\linewidth}
        \centering
        \includegraphics[width=\linewidth]{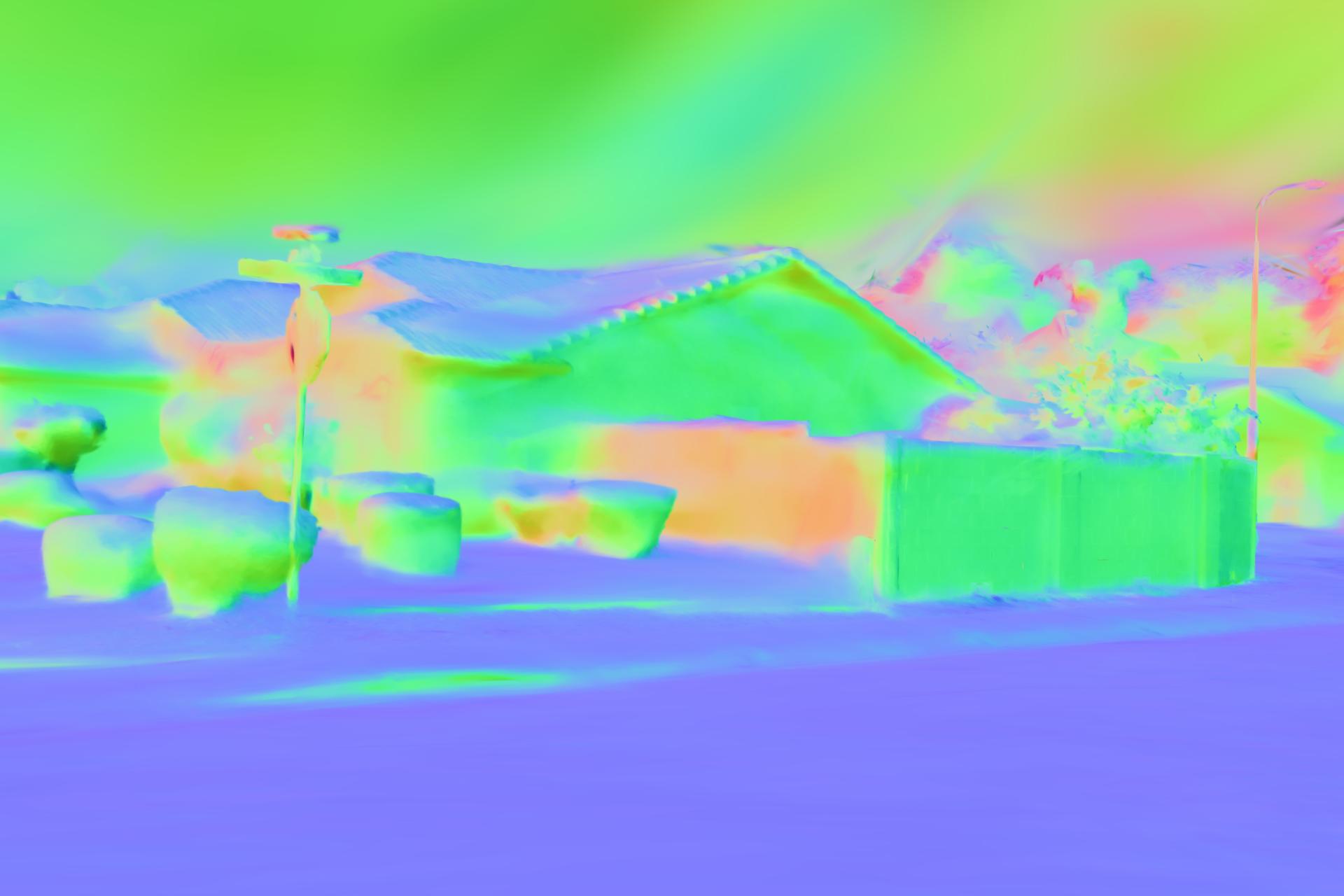}
        \caption[Network2]%
        {{\small Normal map}}    
        \label{fig:mean and std of net14}
    \end{subfigure}
    \begin{subfigure}{0.19\linewidth}
        \centering
        \includegraphics[width=\linewidth]{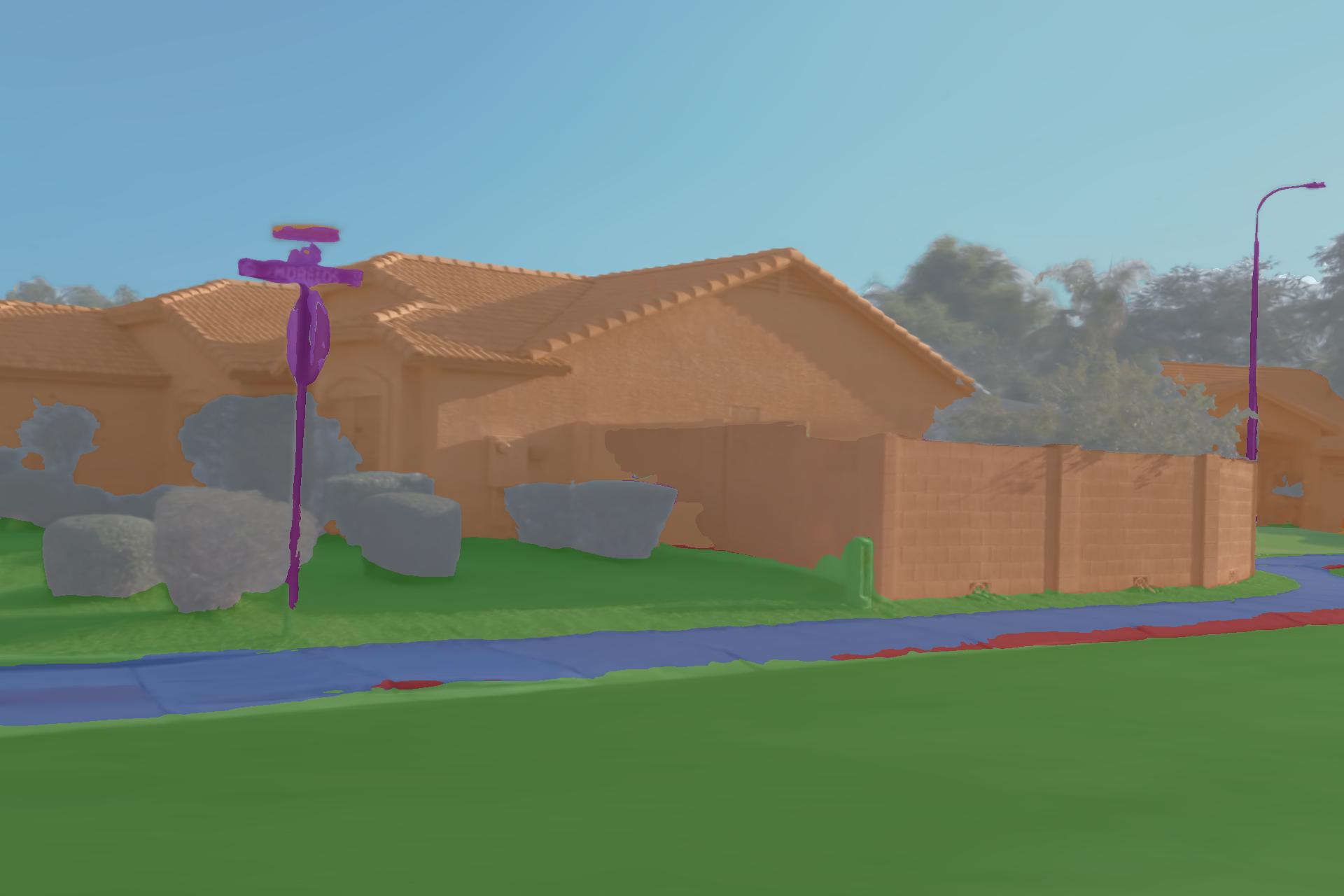}
        \caption[Network2]%
        {{\small Learned segmentation}}    
        \label{fig:mean and std of net14}
    \end{subfigure}
    \caption[ ]
    {\small Novel view multimodal renderings from our GS model.} 
    \label{fig:multimodal}
\end{figure*}

To qualitatively evaluate the effectiveness of our reconstruction method, we visualize its output across multiple modalities. Figure \ref{fig:multimodal} demonstrates that our approach produces high-quality 3D reconstructions, accurately capturing geometric details and surface properties across RGB, depth, LiDAR intensity, normal, and segmentation modalities. Figure \ref{fig:lidar_fig} shows an example of LiDAR point cloud visualization of the hybrid rendering method.

\subsection{Static scene reconstruction quality}
Due to the application of bilateral grid-based color correction in the NeRF step of our reconstruction method, the traditional PSNR metric does not provide a reliable measure of reconstruction quality. Since the SSIM and LPIPS metrics are more sensitive to structural similarity than simple pixel-wise color differences, they are more useful in our case. However, they also remain sensitive to inherent color mismatches between the original and reconstructed images caused by the ISP disentanglement step which is reflected in the evaluation results. In Table \ref{table:SSIM_LPIPS}, we present these metrics on six Waymo scenes compared to vanilla 3DGS and TCLC-GS \cite{zhao2024tclcgstightlycoupledlidarcamera} methods. Beyond these perceptual metrics, we further evaluate our method's efficacy through downstream task performance, which aligns more closely with our objective of generating high-fidelity simulation environments for autonomous vehicle development.

\subsection{Novel view synthesis quality}
% We evaluate the novel view synthesis quality of our method against omniRe \cite{chen2024omnire}, which is considered SOTA model for supervised dynamic scene reconstruction. For quantitative comparison, we trained omniRe and our model on the same scenes from the Waymo dataset, using all five cameras, then we used a set of evaluation images for metric calculation. Since our method aims to reconstruct only the static parts of the scene, we only compute the PSNR, SSIM metrics for the static parts of the scene. In table \ref{table:omnire_compar} we compare our method to omniRe on three dynamic Waymo scene, using the aforementioned metrics. For more implementation details, please refer to the appendix.

We evaluate the novel view synthesis performance of our method in comparison with OmniRe \cite{chen2024omnire}, which is considered the state-of-the-art approach for supervised dynamic scene reconstruction. For quantitative evaluation, both OmniRe and our model were trained on the same scenes from the Waymo dataset using all five cameras, and a separate set of held-out images was used for metric computation. Because our method focuses on reconstructing only the static components of the environment, we report PSNR and SSIM scores computed exclusively over the static regions. Table \ref{table:omnire_compar} presents a comparison of our method and OmniRe across three dynamic Waymo scenes using these metrics. Additional implementation details are provided in the Appendix.

% While evaluating against "left out" images from the original dataset provides quantitative results, this evaluation metric only measures the interpolation capability of the models without providing meaningful information about the robustness of the model for more severe viewpoint deviations from the original train trajectory (e.g. lane changes) which are quite common in autonomous driving related use cases. To address this, we also conducted qualitative novel view synthesis comparisons between omniRe and our model using laterally shifted ego trajectories up to 5 meters.

Although evaluating "left out" images from the original dataset yields quantitative insights, this setup primarily measures the models’ interpolation capabilities and does not adequately reflect their robustness to significant viewpoint shifts relative to the training trajectory (e.g., lane changes), which are common in autonomous driving scenarios. To address this limitation, we also perform qualitative novel view synthesis comparisons between OmniRe and our method under laterally shifted ego trajectories of up to 5 meters.

%Figures \ref{fig:omnire_compar_novel_big} and \ref{fig:omnire_compar_novel} show the qualitative differences from extreme novel views for autonomous driving critical sections such as lane markings, traffic signs, and pedestrian crossings. Our model achieves better quality in these situations due to the explicit normal regularization, which forces our flat splats to align with the surfaces reconstructing not only the rgb images, but the actual geometry of the scene more realistically.

\begin{figure}[b]
    \centering
    \includegraphics[width=1.0\linewidth]{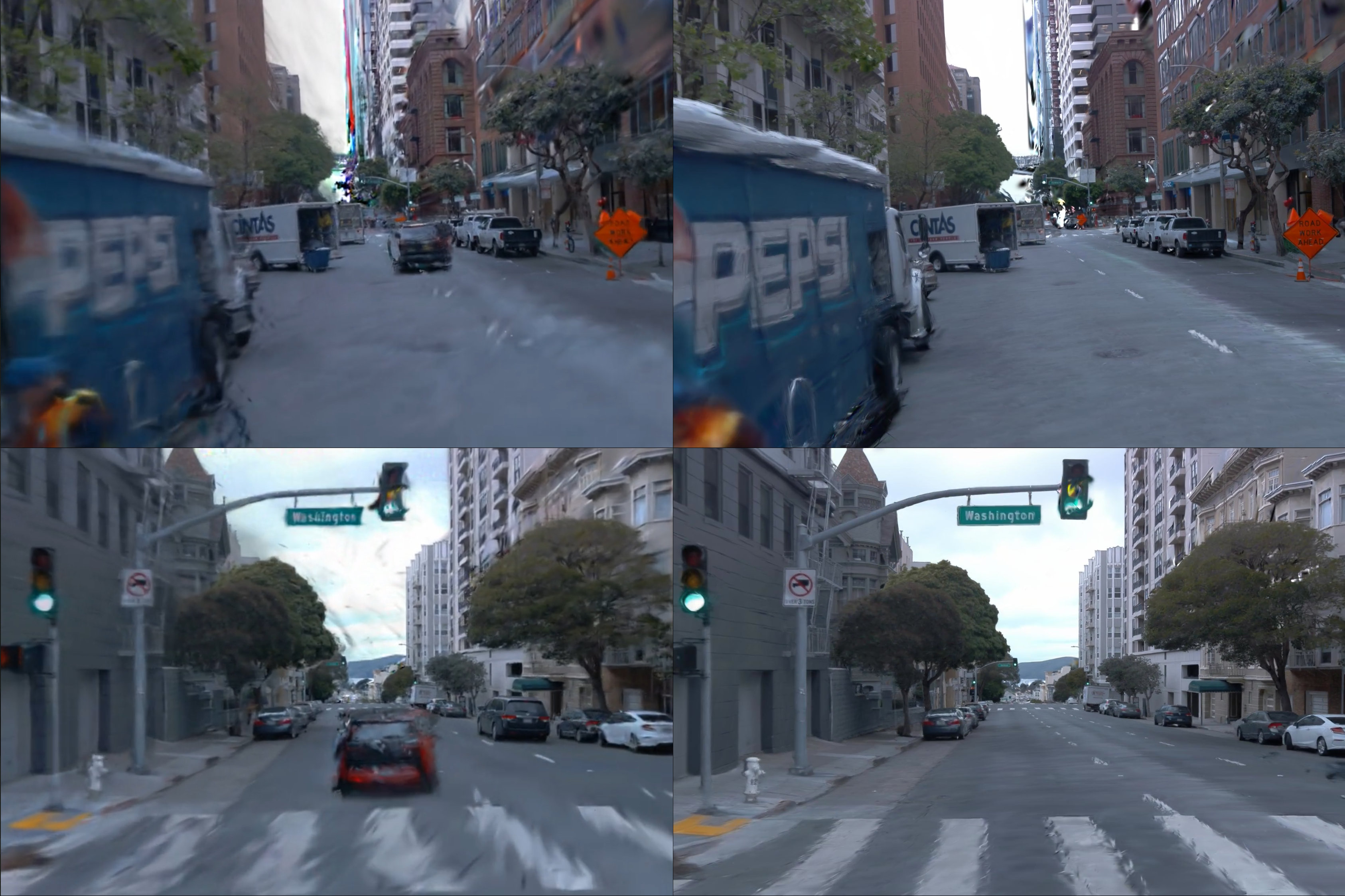}
    \caption{Qualitative comparison of novel view synthesis quality between OmniRe (left) and our model (right).}
    \label{fig:omnire_compar_novel}
\end{figure}

Figures \ref{fig:omnire_compar_novel_big} and \ref{fig:omnire_compar_novel} demonstrate the qualitative differences in highly extrapolated views, focusing on autonomous driving critical features like lane markings, traffic signs, and pedestrian crossings. We assume our model achieves superior quality in these scenarios due to the explicit normal regularization. This regularization constrains the flat splats to align with scene surfaces, leading to a more realistic reconstruction of the actual geometry, in addition to the accurate rendering of RGB images.

% \begin{figure*}
%     \centering
%     \includegraphics[width=1.0\linewidth]{ICCV-author-kit/images/omnire_vs_our_cut_big.jpg}
%     \caption{Qualitative comparison of novel view synthesis quality between omniRe (left) and our model (right) for the static part of the scene.}
%     \label{fig:omnire_compar_novel_big}
% \end{figure*}

\subsection{Downstream tasks}

%\begin{table}[h]
%    \centering
%    \scriptsize
%    \begin{tabular}{|c|c|c|c|c|}
%        \hline
%        \textbf{Sequence} & \makecell{Our 3DGS \\ SSIM$\uparrow$/LPIPS} & \makecell{Our NeRF \\ SSIM$\uparrow$/LPIPS} & \makecell{Vanilla 3DGS$^*$ \\ SSIM$\uparrow$/LPIPS} & \makecell{TCLC-GS$^*$ \\ SSIM$\uparrow$/LPIPS} \\ \hline
%        10247954...    & 0.88/0.26  & 0.88/0.22 &  0.84/0.25  & 0.89/0.16\\ \hline
%        10713922...    & 0.82/0.33  & 0.83/0.27 &  0.84/0.25  & 0.88/0.17\\ \hline
%        11037651...    & 0.75/0.58  & 0.76/0.53  &  0.67/0.44 & 0.71/0.42\\ \hline
%        13469905...    & 0.87/0.26  & 0.88/0.21  &  0.84/0.26 & 0.88/0.16\\ \hline
%        14333744...    & 0.85/0.28  & 0.86/0.24  &  0.85/0.25 & 0.87/0.21\\ \hline
%        14663356...    & 0.84/0.38  & 0.84/0.31  &  0.86/0.25 & 0.90/0.18\\ \hline
%    \end{tabular}
%    \caption{Reconstruction quality comparison on Waymo scenes. $^*$marked results are taken from the TLCL-GS \cite{zhao2024tclcgstightlycoupledlidarcamera} paper.}
%    \label{table:SSIM_LPIPS}
%\end{table}

\begin{table}[t]
    \centering
    \scriptsize
    \begin{tabular}{|c|c|c|c|}
        \hline
        \textbf{Sequence} & \makecell{\textbf{Our 3DGS} \\ \textbf{SSIM}$\uparrow$\textbf{/LPIPS}$\downarrow$} & \makecell{\textbf{Vanilla 3DGS$^*$} \\ \textbf{SSIM}$\uparrow$/\textbf{LPIPS}$\downarrow$} & \makecell{\textbf{TCLC-GS$^*$} \\ \textbf{SSIM}$\uparrow$/\textbf{LPIPS}$\downarrow$} \\ \hline
        10247954...    & 0.88/0.26  &  0.84/0.25  & 0.89/0.16\\ \hline
        10713922...    & 0.82/0.33  &  0.84/0.25  & 0.88/0.17\\ \hline
        11037651...    & 0.75/0.58  &  0.67/0.44 & 0.71/0.42\\ \hline
        13469905...    & 0.87/0.26  &  0.84/0.26 & 0.88/0.16\\ \hline
        14333744...    & 0.85/0.28  &  0.85/0.25 & 0.87/0.21\\ \hline
        14663356...    & 0.84/0.38  &  0.86/0.25 & 0.90/0.18\\ \hline
    \end{tabular}
    \caption{Reconstruction quality comparison on Waymo scenes. $^*$marked results are taken from the TLCL-GS \cite{zhao2024tclcgstightlycoupledlidarcamera} paper.}
    \label{table:SSIM_LPIPS}
\end{table}

\begin{table}[t]
    \centering
    \scriptsize
    \begin{tabular}{|c|c|c|}
        \hline
        \textbf{Sequence} & \makecell{\textbf{Our 3DGS} \\ \textbf{PSNR}$\uparrow$\textbf{SSIM}$\uparrow$} & \makecell{\textbf{OmniRe} \\ \textbf{PSNR}$\uparrow$/\textbf{SSIM}$\uparrow$} \\ \hline
        13299463...    & 26.15/0.8011  &  26.07/0.7908\\ \hline
        17860546...    & 30.13/0.9017  &  29.89/0.8894\\ \hline
        30154365...    & 28.36/0.8767  &  27.62/0.8433\\ \hline
    \end{tabular}
    \caption{Novel view synthesis quality comparison on Waymo scenes.}
    \label{table:omnire_compar}
\end{table}

\begin{figure}[b]
    \centering
    \includegraphics[width=1.0\linewidth]{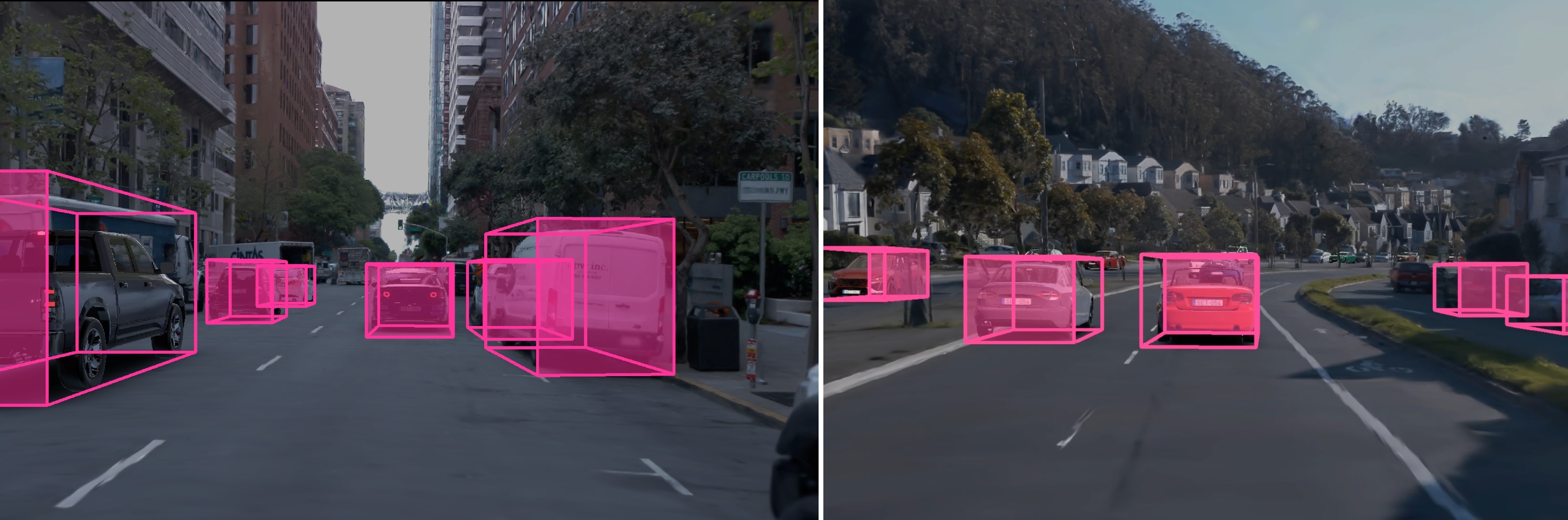}
    \caption{DEVIANT model detections on images rendered using our hybrid method.}
    \label{fig:detections_3dod}
\end{figure}

\begin{table*}[t]
    \centering
    \renewcommand{\arraystretch}{0.9}
    \begin{tabular}{c | c | c|ccc | c|ccc}
        \toprule
        \multirow{2}{*}{\textbf{IoU$_{3D}$}} & \multirow{2}{*}{\textbf{Scenario}} & \multicolumn{4}{c}{\textbf{AP$_{3D}$ [\%] $\uparrow$}} & \multicolumn{4}{c}{\textbf{APH$_{3D}$ [\%] $\uparrow$}} \\
        \cmidrule(lr){3-6} \cmidrule(lr){7-10}
        &  & All & 0-30 & 30-50 & 50-$\infty$ & All & 0-30 & 30-50 & 50-$\infty$ \\
        \midrule
        \multirow{5}{*}{0.3}
        &   Original    & 11.08 & 20.16 & 23.42 & 0.54 & 10.99 & 20.04 & 23.00 & 0.53 \\
        & Shifted (0 m) &  9.40 & 23.88 & 13.73 & 0.39 &  9.34 & 23.73 & 13.55 & 0.38 \\
        & Shifted (1 m) &  8.17 & 22.57 & 11.46 & 0.00 &  8.08 & 22.46 & 10.91 & 0.00 \\
        & Shifted (2 m) &  9.17 & 22.38 & 15.04 & 0.00 &  9.02 & 21.76 & 14.86 & 0.00 \\
        & Shifted (3 m) &  9.93 & 22.69 & 14.75 & 0.00 &  9.86 & 22.48 & 14.56 & 0.00 \\        
        \bottomrule
    \end{tabular}
    \caption{Comparison of 3D object detection results on a scenario rendered from novel views.}
    \label{tab:results_3dod}
\end{table*}

\begin{table}[t]
    \centering
    \footnotesize
    \renewcommand{\arraystretch}{0.9}
    \begin{tabular}{c | cc | cc | cc}
        \toprule
        \multirow{2}{*}{\textbf{IoU$_{3D}$}} & \multicolumn{2}{c}{\textbf{AP$_{3D}$ [\%] $\uparrow$}} & \multicolumn{2}{c}{\textbf{APH$_{3D}$ [\%] $\uparrow$}} & \multicolumn{2}{c}{\textbf{Recall@0.95}} \\
        \cmidrule(lr){2-3} \cmidrule(lr){4-5} \cmidrule(lr){6-7}
        & 0–30 & 30–50 & 0–30 & 30–50 & 0–30 & 30–50 \\
        \midrule
        0.5 & 100.00 & 52.94 & 99.05 & 52.09 & 100.00 & 0.00 \\ 
        0.7 & 66.67 & 40.00 & 65.92 & 39.87 & 0.00 & 0.00 \\ 
        \bottomrule
    \end{tabular}
    \caption{Comparison of 3D object detection results using the Detection Agreement metric.}
    \label{tab:results_3dod_da}
\end{table}

\paragraph{3D object detection.} To validate the effectiveness of our solution for 3D object detection, we used DEVIANT \cite{kumar2022deviant}, a publicly available monocular 3D object detection model trained on the Waymo Open Dataset. We conducted both quantitative and qualitative experiments. For the quantitative analysis, we rendered the scenes from a novel viewpoint, shifting the ego-trajectory laterally by varying offsets (ranging from 0 to 3 meters) from \textit{segment-13469905891836363794}, and used the transformed ground truth bounding boxes for metric calculation. The qualitative experiments involved running the DEVIANT model on three segments, where our reconstruction model rendered the static environment, and the mesh-based rendering engine introduced dynamic vehicles. Our primary focus was to assess whether simulator-generated vehicles introduce a domain gap.

Table \ref{tab:results_3dod} presents the detection results using Waymo detection metrics, where the 'Original' scenario refers to using the original images, while 'Shifted' scenarios are generated by our reconstruction model with laterally offset ego-trajectories. Interestingly, the model performs even better in novel view scenarios than on the original images at close range (0–30m), primarily due to a higher number of false positives in the original images. Distant object detection performance declines in novel view synthesis, warranting further investigation on a larger sample. However, detection results remain largely consistent across novel view scenarios. The observed differences can be attributed to object visibility changes caused by lateral shifts, as the model struggles to recognize truncated objects. In addition, we evaluated the detection results on the original and simulated images using the Detection Agreement metric \cite{manivasagam2023towards} (see Table \ref{tab:results_3dod_da}). The strong agreement between detections on original and simulated images suggests a small domain gap. 

% While the dataset is limited in size (5856 GT boxes), we believe these results provide convincing preliminary evidence of our method’s effectiveness.

Qualitative results show that the model successfully detects vehicles rendered by both the reconstruction model and the mesh-based rendering engine, indicating that no significant domain gap is introduced. However, distant objects remain undetected due to the model's detection range limitation. Example detections are shown in Figure \ref{fig:detections_3dod}. 

Additional figures showing the model's detections on data rendered by our hybrid solution, as well as quantitative results of another downstream task (semantic segmentation), can be found in the Supplementary Material.

\section{Limitations and Future Work}

\paragraph{Limitations.} 
The proposed block-based rendering approach, while designed to minimize discontinuities at block boundaries through sufficient overlap, can still produce visible artifacts, particularly in regions such as the sky and highly textured distant objects. Moreover, while RGB sky reconstruction is generally satisfactory, the sky geometry is often placed excessively close to the scene, leading to artifacts in extremely novel viewpoints and depth maps. Additionally, the current representation of LiDAR intensities as static, view-independent values for each Gaussian does not accurately reflect the inherent directional dependence observed in real LiDAR sensors. While the evaluation of autonomous driving downstream tasks yielded promising results, the sample size remains limited, and a larger dataset is needed to draw definitive conclusions.

\paragraph{Future work.} Future research endeavors will primarily focus on addressing the aforementioned limitations and enhancing the realism of rendered scenes, thereby further reducing the domain gap between rendered and recorded data.  Specifically, we intend to explore hierarchical Gaussian Splatting to mitigate block-related artifacts. Furthermore, by leveraging the existing segmentation maps, we aim to decompose scenes into separate sky and ground models. These models will jointly be trained to ensure high-quality reconstruction while preserving the geometric accuracy of the sky-ground separation. Utilizing the generated normal information, we will investigate methods to improve LiDAR intensity simulation, better reflecting the directional dependencies observed in real LiDAR data.

Furthermore, we will investigate the possible improvement in the integration of mesh-based objects into Gaussian Splatting environments to enable realistic interaction modeling. This includes the development of techniques for physically accurate shadow casting, dynamic illumination from inserted light sources (e.g., car headlights), and global relighting. These advancements will facilitate the simulation of diverse environmental conditions, such as varying weather patterns, within the reconstructed scenes.

%%%%%%%%%%%%%%%%%%%%%%%%%%%%%%%%%%%%%%%%%%%%%%%%%%%%%%%%%%%%%%%
% TODO: DO NOT FORGET TO REMOVE FROM THE CAMERA-READY VERSION
%%%%%%%%%%%%%%%%%%%%%%%%%%%%%%%%%%%%%%%%%%%%%%%%%%%%%%%%%%%%%%%
%\iffalse
\section*{Acknowledgment}
This research was conducted using technologies developed at \href{https://aimotive.com/}{aiMotive}, specifically components of the \href{https://aimotive.com/aisim}{aiSim} and \\\href{https://aimotive.com/aidata}{aiData} toolchain. aiSim is a high-fidelity, real-time simulation platform designed for the development and validation of automated driving systems, supporting both traditional and hybrid rendering to achieve sensor-accurate, scalable, and deterministic testing. aiData is aiMotive’s data management and processing solution, providing high-quality, diverse datasets and automated annotation pipelines to accelerate AI model training and validation. These tools are commercially available and have been widely adopted to support the safe and efficient development of advanced driver-assistance and automated driving systems.
%\fi
%\input{sec/2_formatting}
%\input{sec/3_finalcopy}
{
    \small
    \bibliographystyle{ieeenat_fullname}
    \bibliography{main}
}

%%%%%%%%%%%%%%%%%%%%%%%%%%%%%%%%%%%%%%%%%%%%%%%%%%%%%%%%%%%
\newpage
\section*{\textbf{\LARGE Supplementary Material}}
\appendix

\section{Quantitative results}

\subsection{Novel view synthesis quality comparisons}
In addition to comparing against OmniRe \cite{chen2024omnire}, we further evaluated the novel view synthesis performance of our method relative to DeSiRe-GS \cite{peng2024desire}, a state-of-the-art unsupervised dynamic scene reconstruction approach for urban environments. Table \ref{table:desire_compar} reports the PSNR and SSIM scores for DeSiRe-GS alongside those of OmniRe and our method. In the case of OmniRe and DeSiRe-GS, we used their official repositories for training and metrics calculations. Across the evaluated scenarios, our approach matches or exceeds the performance of existing state-of-the-art neural reconstruction techniques. Note that our method uses a bilateral guided ISP disentanglement method. While this approach mitigates camera appearance inconsistencies (which are highly relevant in real-world applications), it also has a negative impact on our PSNR and SSIM scores.

\begin{table}[b]
    \centering
    \scriptsize
    \begin{tabular}{|c|c|c|c|}
        \hline
        \textbf{Sequence} & \textbf{Our 3DGS} & \textbf{OmniRe} & \textbf{DeSiRe-GS} \\
        & \textbf{PSNR$\uparrow$/SSIM$\uparrow$} & \textbf{PSNR$\uparrow$/SSIM$\uparrow$} & \textbf{PSNR$\uparrow$/SSIM$\uparrow$} \\ \hline
        13299463... & \textit{26.15/0.8011} & 26.07/0.7908 & \textbf{26.94/0.8071}\\ \hline
        17860546... & \textbf{30.13/0.9017} & \textit{29.89/0.8894} & 28.58/0.8499\\ \hline
        30154365... & \textbf{28.36/0.8767} & \textit{27.62/0.8433} & 26.03/0.7908\\ \hline
    \end{tabular}
    \caption{Novel view synthesis quality comparison on Waymo scenes. The highest-performing method is highlighted in \textbf{bold}, while the second-best result is indicated in \textit{italics}.}
    \label{table:desire_compar}
\end{table}

\subsection{Downstream task: Semantic Segmentation}
Our evaluation leverages semantic segmentation to analyze model performance in three key areas. Firstly, we evaluate the accuracy of our 3D reconstructions. This is achieved by comparing segmentation masks produced by Mask2Former \cite{cheng2021mask2former} on ground truth images with those generated from renderings of our NeRF and 3DGS models. To ensure a fair comparison, ground truth images are reprojected to align with the undistorted pinhole camera parameters used for rendering. We quantify the reconstruction quality by calculating the percentage of pixels with matching classifications. To mitigate potential bias from easily segmented regions like road surfaces and sky, we also report the Intersection over Union (IoU) specifically for the 'CAR' class. It is important to note that this metric also incorporates errors from the Mask2Former model, for example in cases where our segmentation accurately predicts the labels for distant/obstructed objects but Mask2Former fails to do so, based on the more limited information.

Our evaluation employs semantic segmentation to assess model performance across three key aspects. First, we measure the accuracy of our 3D reconstructions by comparing segmentation masks generated by Mask2Former \cite{cheng2021mask2former} on ground truth images with those derived from renderings of our NeRF and 3DGS models. To ensure a fair comparison, we reproject the ground truth images to match the undistorted pinhole camera parameters used for rendering. Reconstruction quality is quantified by calculating the percentage of pixels with matching classifications. To reduce potential bias from easily segmented regions such as roads and sky, we also report the Intersection over Union (IoU) specifically for the 'CAR' class. It is important to note that this metric includes errors from the Mask2Former model—for instance, in cases where our segmentation correctly labels distant or obstructed objects, but Mask2Former fails due to its limited visual information. As shown in Table \ref{table:segm1}, the segmentation masks generated from ground truth images exhibit a high degree of overlap with those produced from the neural reconstruction, with two exceptions. The lower-performing cases correspond to a nighttime recording and a scene where a camera is pointed directly at the sun.

\begin{table}[h]
    \centering
    \begin{tabular}{|c|c|c|c|}
        \hline
        \textbf{Sequence} & \textbf{GT-3DGS} & \textbf{GT-NeRF} & \textbf{GT-3DGS (Car)} \\ \hline
        10247954...    & 97.0\%     & 97.52\% &  96.7\%   \\ \hline
        10713922...    & 93.5\%     & 94.4\%  &  87.6\%  \\ \hline
        11037651...    & 83.5\%     & 86.8\%  &  60.3\%  \\ \hline
        13469905...    & 93.9\%     & 95.4\%  &  95.5\%  \\ \hline
        14333744...    & 94.4\%     & 95.4\%  &  94.8\%  \\ \hline
        14663356...    & 86.3\%     & 89.3\%  &  78.4\%  \\ \hline
    \end{tabular}
    \caption{Percentage of pixels classified the same between the ground truth and 3DGS/NeRF rendered images measured on Waymo scenes.}
    \label{table:segm1}
\end{table}

\begin{figure}
    \centering
    \includegraphics[width=0.48\linewidth]{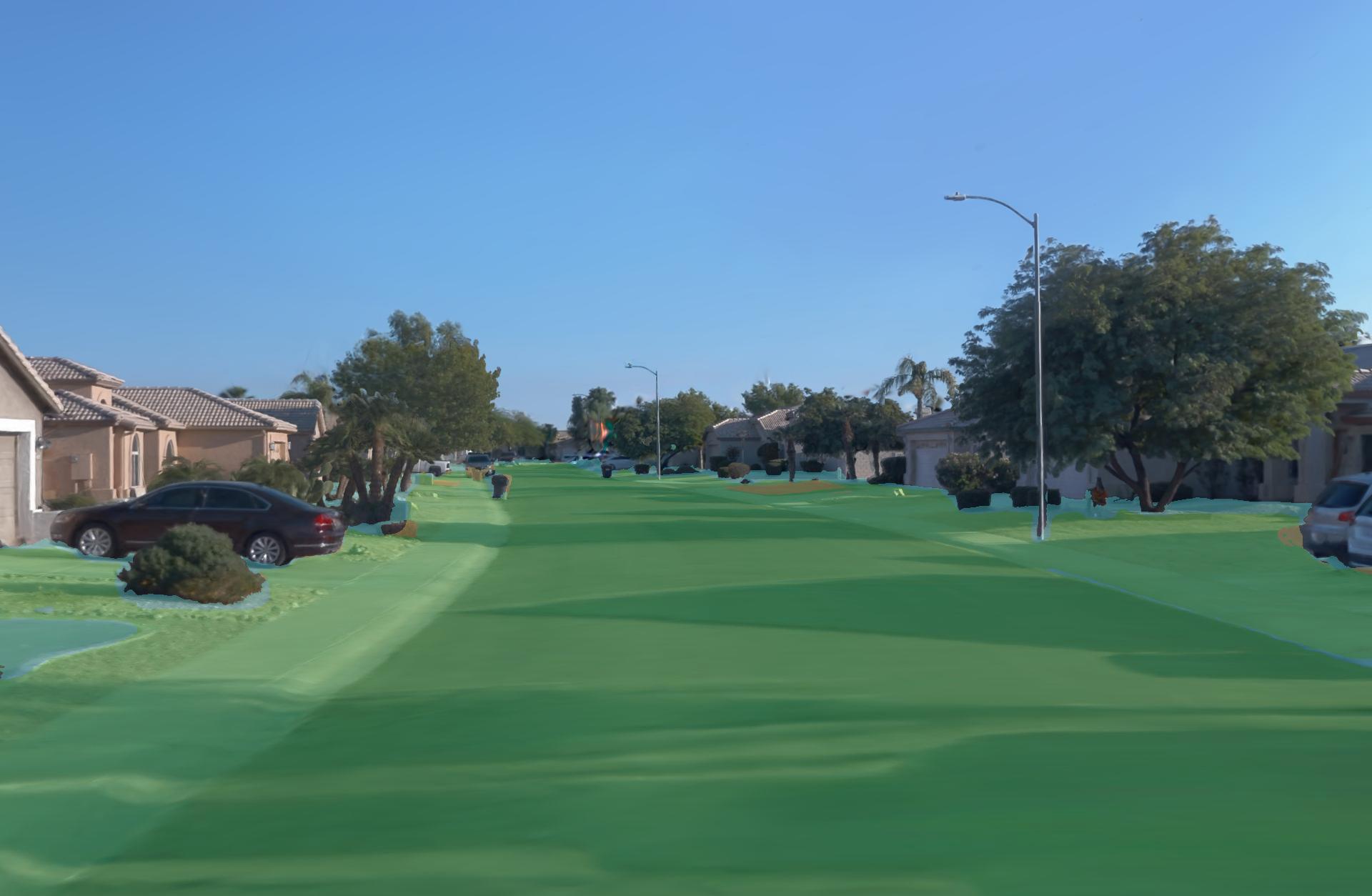}
    \includegraphics[width=0.48\linewidth]{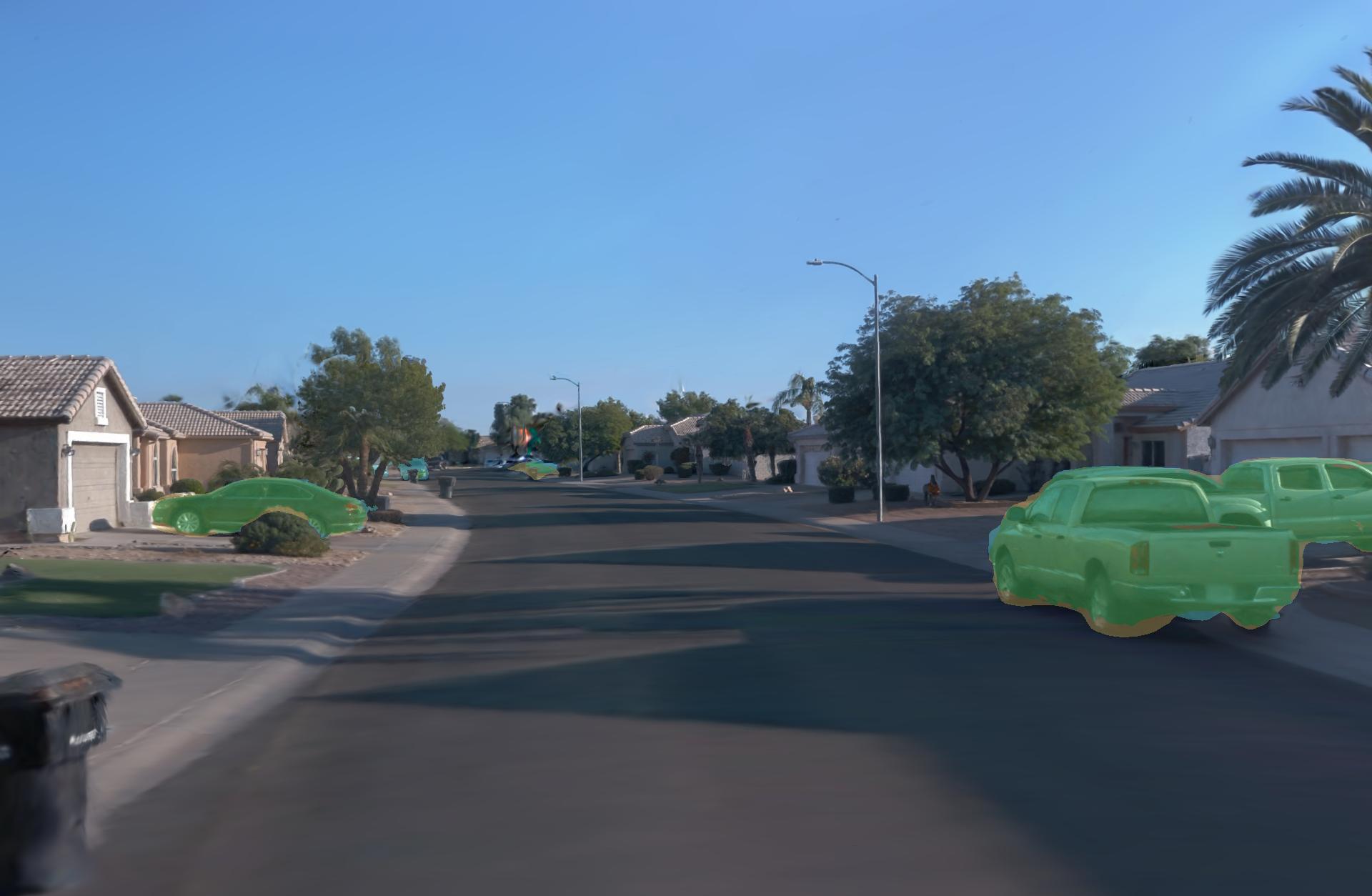}
    \caption{Comparison of learned segmentation with the Mask2Former model. Green regions correspond to matches between the two models, while blue and orange colors represent regions classified by only the learned segmentation or Mask2Former, respectively.}
    \label{fig:segm_comparison}
\end{figure}

Secondly, we assess the joint performance of novel view synthesis and learned segmentation. We render images from unseen viewpoints and compare the segmentation masks generated by Mask2Former on these novel views to our model's predicted segmentation outputs, thus evaluating both aspects simultaneously. Novel views were generated along the ego-vehicle's trajectory, with horizontal displacements ranging from -1m to 3m, using a forward-facing virtual camera. We evaluate performance on the 'CAR' class and a combined 'road-like' class, which encompasses 'ROAD', 'SIDEWALK', 'PARKING', and 'LANE MARKING', to align with our internal segmentation model's labeling. The results presented in Table \ref{table:segm_comparison} and in Figure \ref{fig:segm_comparison} indicate robust reconstruction of road features, while the comparatively lower performance for the 'CAR' label is partially attributable to Mask2Former's inability to detect distant or partially obstructed vehicles, thereby amplifying the discrepancy between its predictions and our learned segmentations.

\begin{table}[h]
    \centering
    \begin{tabular}{|c|c|c|}
        \hline
        \textbf{Displacement (m)} & \textbf{IoU (Car)} $\uparrow$ & \textbf{IoU (Roadlike)} $\uparrow$ \\
        \hline
        -1   & 86.5\%   & 93.5\%   \\
        \hline
        0   & 83.4\%   & 94.6\%   \\
        \hline
        1   & 81.5\%   & 94.5\%   \\
        \hline
        2  & 79.8\%  & 94.4\%  \\
        \hline
        3  & 78.6\%  & 93.5\%  \\
        \hline
    \end{tabular}
    \caption{IoU values between Mask2Former and learned segmentation of different classes on novel viewpoint renderings.}
    \label{table:segm_comparison}
\end{table}

\begin{figure}
    \centering
    \includegraphics[width=0.48\linewidth]{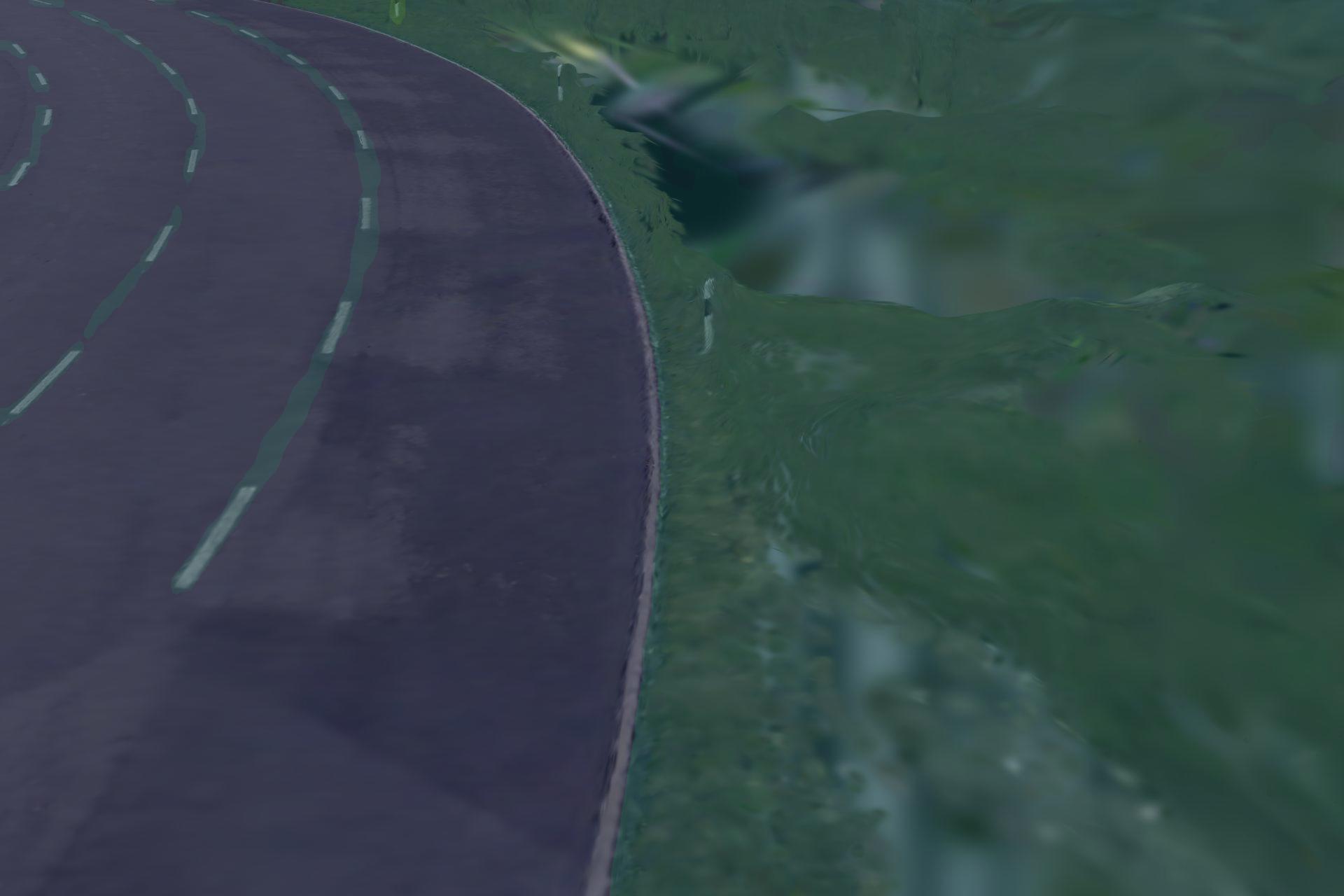}
    \includegraphics[width=0.48\linewidth]{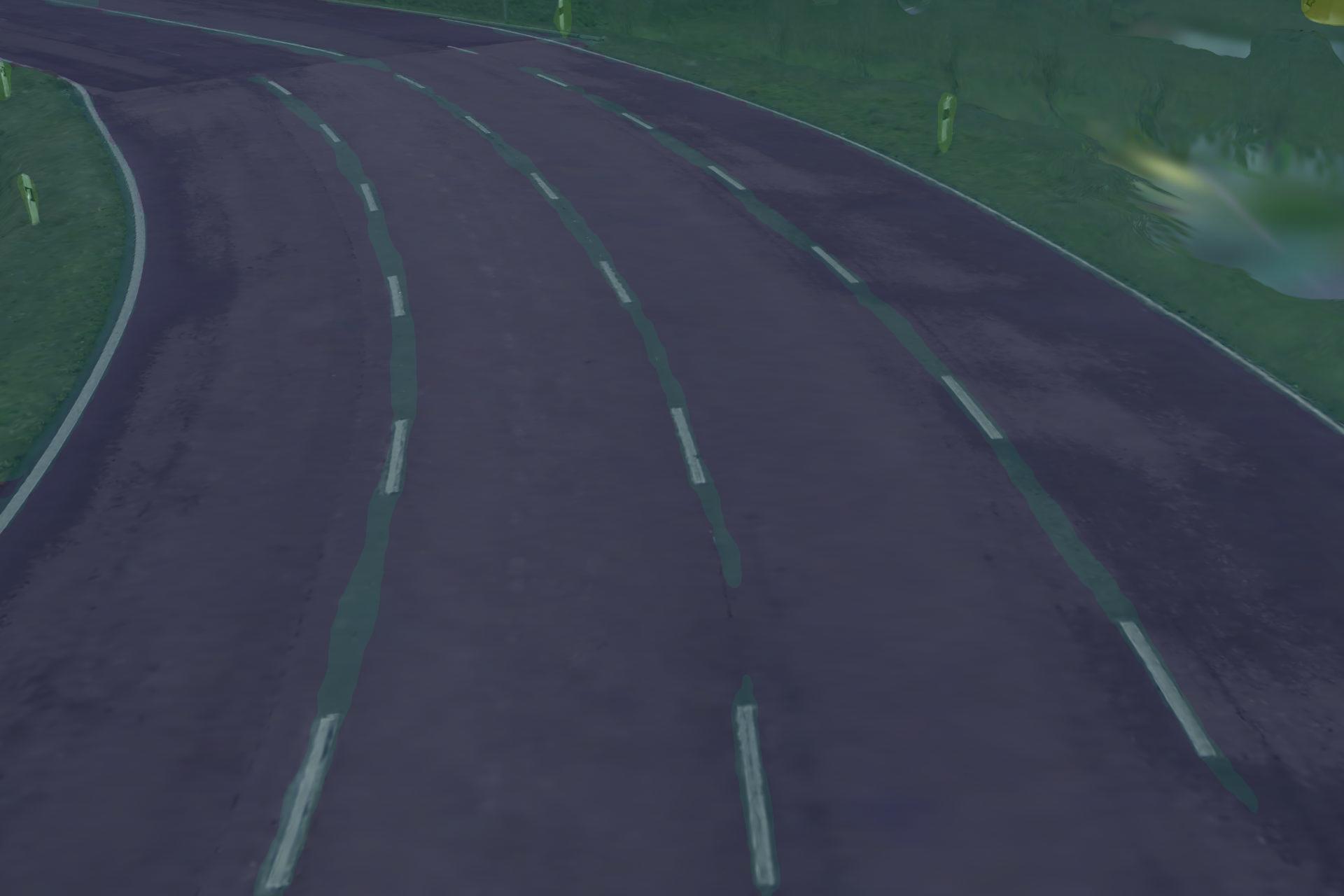}
    \caption{Segmentation of road features using a Mask2Former model from extremely novel viewpoint renderings.}
    \label{fig:ZZ_down}
\end{figure}

\section{Qualitative results}

\subsection{Nerf-based reconstruction}
NeRF renders containing RGB, depth, normal and segmentation outputs are provided in the "Nerf\_render" folder. The videos are rendered using $180^\circ$ horizontal vertical field of view equirectangular cameras on evaluation trajectories.

\subsection{3DGS reconstruction}
3DGS renders rendered using significantly modified camera trajectories on a large-scale dataset can be found in the "3DGS\_render" demonstrating the extrapolation capabilities of our model.

\subsection{Hybrid rendering}
Scenes containing dynamic actors, rendered using our complete hybrid rendering solution can be found in the "hybrid\_render\_raster" and "hybrid\_render\_raytrace" folders. The videos were rendered on modified trajectories using non-pinhole camera models (omnidirectional, and equirectangular), with our rasterization-based, and ray tracing rendering engines.

\subsection{Extreme novel view rendering}
\begin{figure}[h]
  \centering
   \includegraphics[width=1.0\linewidth]{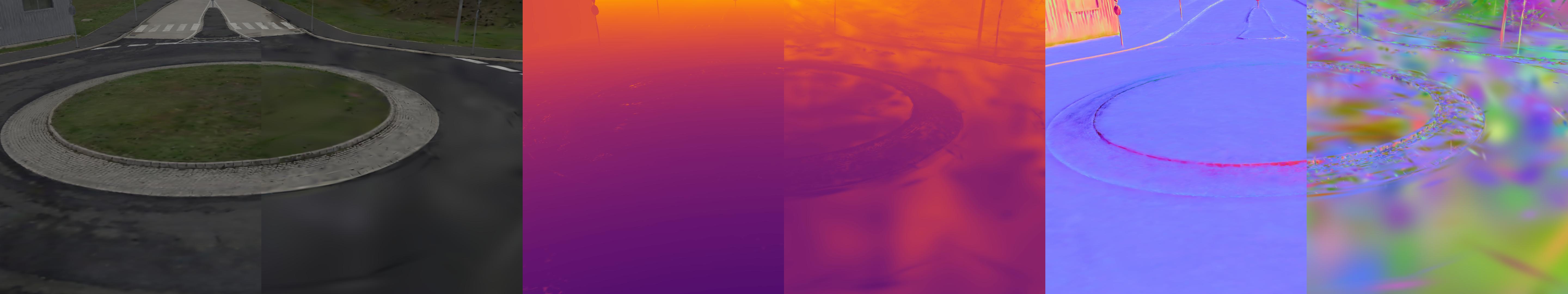}
   %\caption{RGB, depth, and normal images from our (left) and vanilla GS (right) models from extreme novel viewpoint}
   \caption{RGB, depth, and normal images from our (left) and vanilla GS (right) models from extreme novel viewpoint}
   \label{fig:compare}
\end{figure}

\begin{figure*}
    \centering
    \begin{subfigure}{1.0\linewidth}
        \centering
        \includegraphics[width=1.0\linewidth]{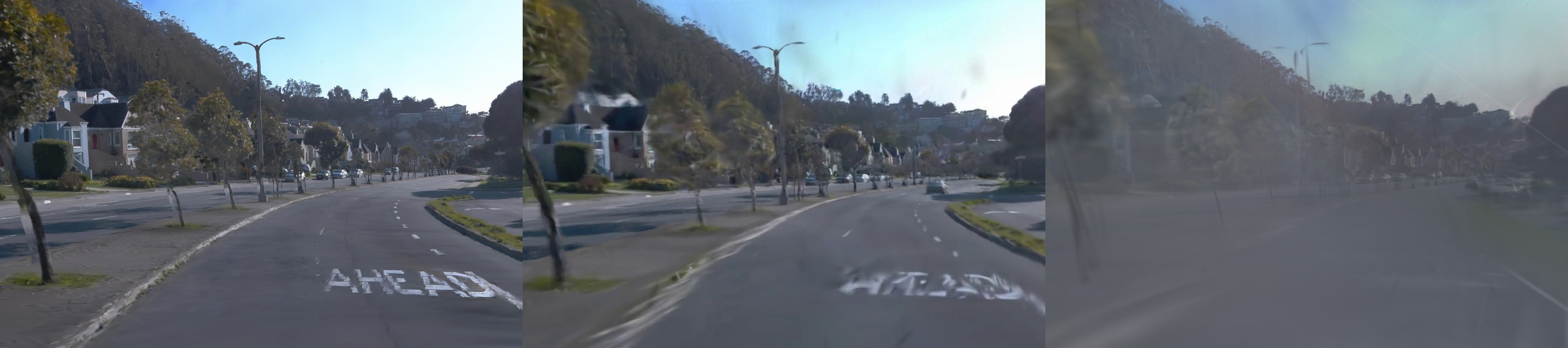}
        \caption{Novel view images shifted by 3 meters from the ego trajectory on the Waymo segment-132... scene}
        \label{fig:132_novel}
    \end{subfigure}
    
    \vspace{0.5cm} % Adds vertical space between the figures

    \begin{subfigure}{1.0\linewidth}
        \centering
        \includegraphics[width=1.0\linewidth]{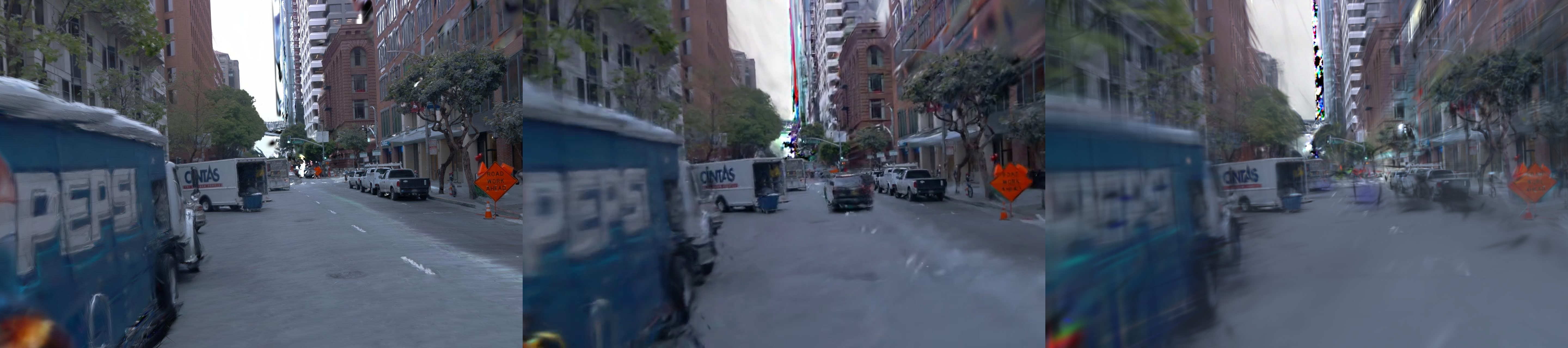}
        \caption{Novel view images shifted by 5 meters from the ego trajectory on the Waymo segment-178... scene}
        \label{fig:178_novel}
    \end{subfigure}
    
    \vspace{0.5cm} % Adds vertical space between the figures

    \begin{subfigure}{1.0\linewidth}
        \centering
        \includegraphics[width=1.0\linewidth]{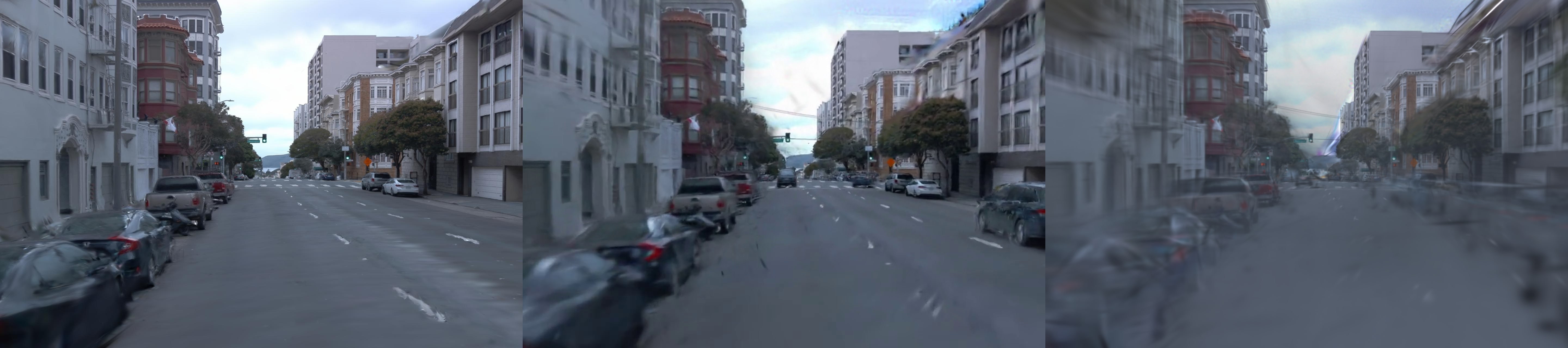}
        \caption{Novel view images shifted by 4 meters from the ego trajectory on the Waymo segment-301... scene}
        \label{fig:301_novel}
    \end{subfigure}
    
    \vspace{0.5cm} % Adds vertical space between the figures
    
    \begin{subfigure}{1.0\linewidth}
        \centering
        \includegraphics[width=1.0\linewidth]{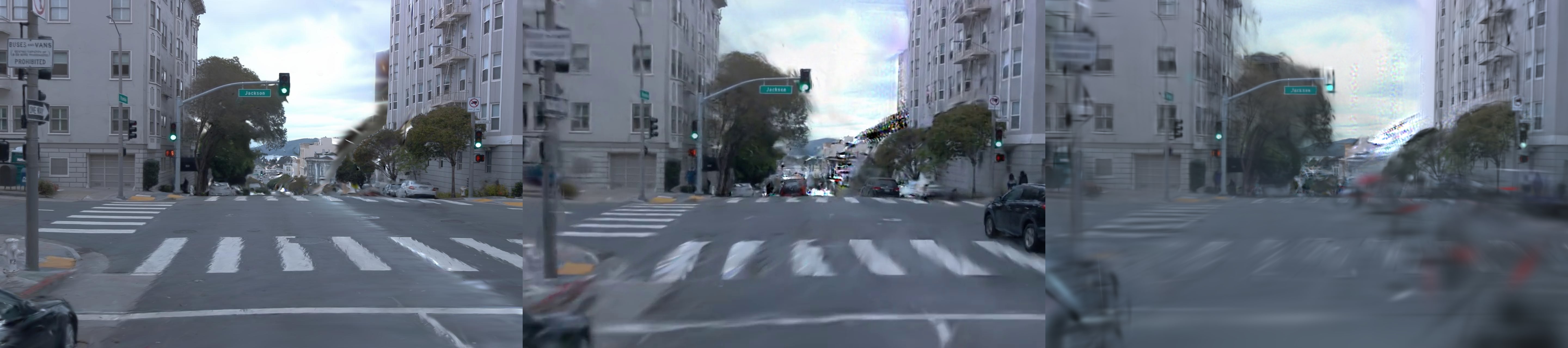}
        \caption{Novel view images shifted by 4 meters from the ego trajectory on the Waymo segment-301... scene}
        \label{fig:301_novel_v2}
    \end{subfigure}
    
    \caption{Extreme novel view renders with our model, OmniRe, and DeSiRe-GS, respectively}
    \label{fig:extreme_novel_combined}
\end{figure*}

On Figure $\ref{fig:compare}$, we illustrate the typical challenges of vanilla 3D Gaussian Splatting when it comes to untextured surfaces and handling extremely novel views. The resulting depth maps exhibit large errors because using only the RGB images does not contain sufficient information for accurate geometry reconstruction, and without direct regularization, the normals of the Gaussians are ill-defined. This failure of the base method significantly hinders its usefulness in autonomous driving simulation and  motivates the need for the specialized approaches evaluated in the following section.
Figure $\ref{fig:extreme_novel_combined}$ shows novel view renders generated from poses laterally shifted by up to $5\text{m}$ from the original trajectory, comparing the performance of DeSiRe-GS, OmniRe, and the method presented herein. Since ground-truth imagery is unavailable for these substantial spatial shifts, quantitative error metrics cannot be computed. The comparison thus relies on a qualitative analysis centered on features deemed crucial for autonomous vehicle perception, including lane markings, pedestrian crossings, traffic signs, and vehicle shapes. Across the examined Waymo scenes, DeSiRe-GS frequently generates visually unacceptable outputs characterized by significant blurring and floaters. Both OmniRe and our model, however, produce coherent renderings. The performance divergence between OmniRe and our approach is most pronounced when examining these perception-critical elements. Specifically, OmniRe introduces distorted lane markings and crosswalks and blurry traffic signs. In contrast, our proposed model consistently maintains superior fidelity and geometric integrity of these essential road features. For a dynamic assessment of these results, we refer the reader to the supplementary comparison videos.

\subsection{Downstream task: Semantic Segmentation}

\begin{figure*}
    \centering
    \includegraphics[width=1.0\linewidth]{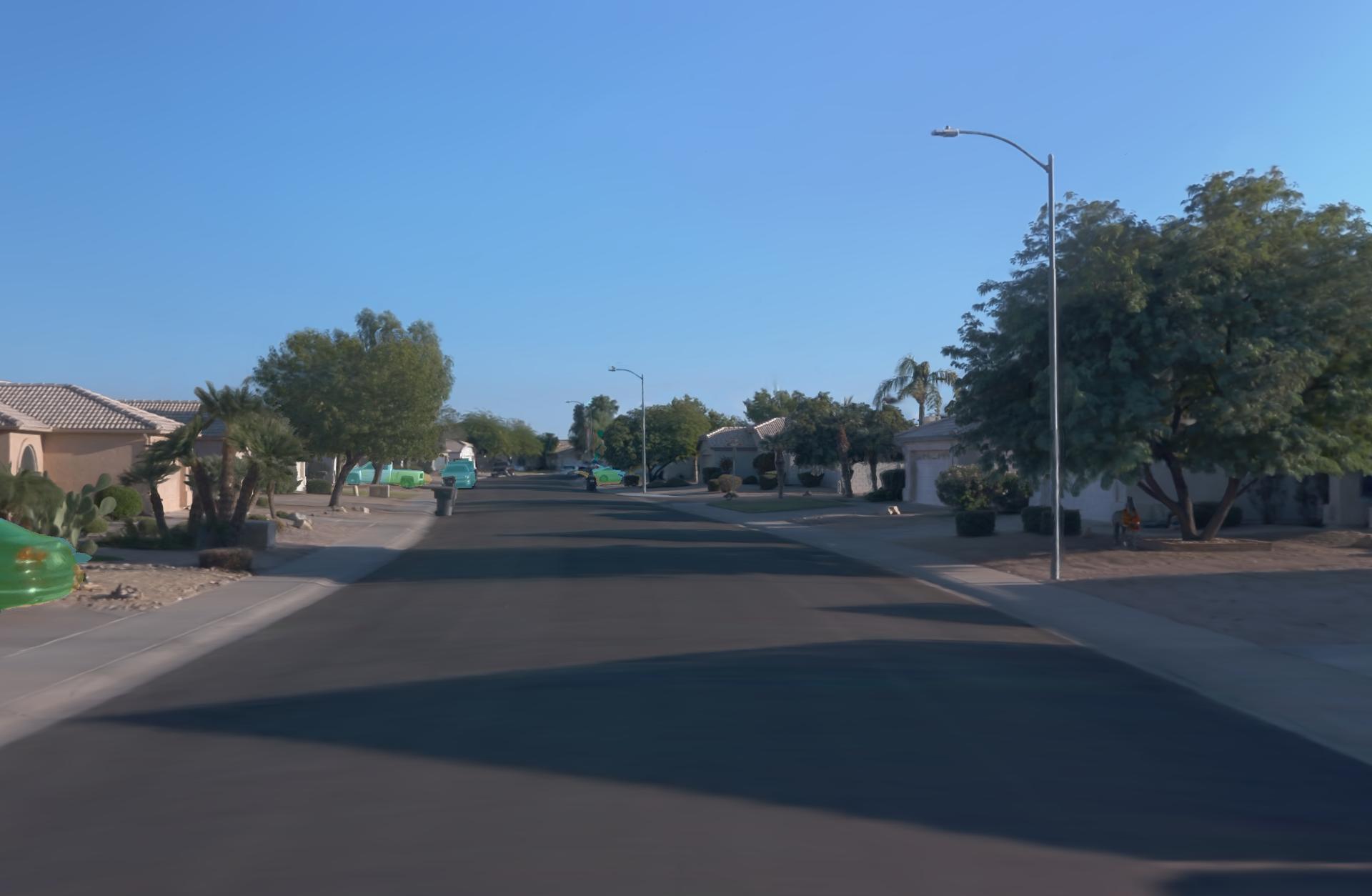}
    \caption{Comparison of 'CAR' label detection between Mask2Former prediction and our learned segmentation, blue regions correspond to parts where our learned segmentation predicted 'CAR' label while Mask2Former did not.}
    \label{fig:car_detection_diff}
\end{figure*}

Figure \ref{fig:car_detection_diff} illustrates a comparison between our learned segmentations and Mask2Former's results for the 'CAR' label. Regions marked in blue are classified as 'CAR' by our learned segmentation model, but not by Mask2Former. Our approach achieves accurate segmentations, even for partially obstructed or distant objects, due to the association of class labels with discrete 3D object representations (Gaussians). These representations are trained from diverse viewpoints and distances, rather than relying solely on 2D pixel information. Consequently, this example demonstrates that the comparison metric presented in the main text may underestimate our actual performance, as it incorporates Mask2Former's erroneous predictions.
We also conducted a qualitative evaluation of our extreme novel view renderings of the road surface by employing a segmentation network to detect lanes, roadsides, and the road itself. As depicted in Figure \ref{fig:ZZ_down}, the network successfully detected these critical features in images rendered from a 5-meter height, laterally shifted by 2 meter, and with the camera pitched down by 25°.

\subsection{Downstream task: 3D Object Detection}
\begin{figure*}
    \centering
    \includegraphics[width=1.0\linewidth]{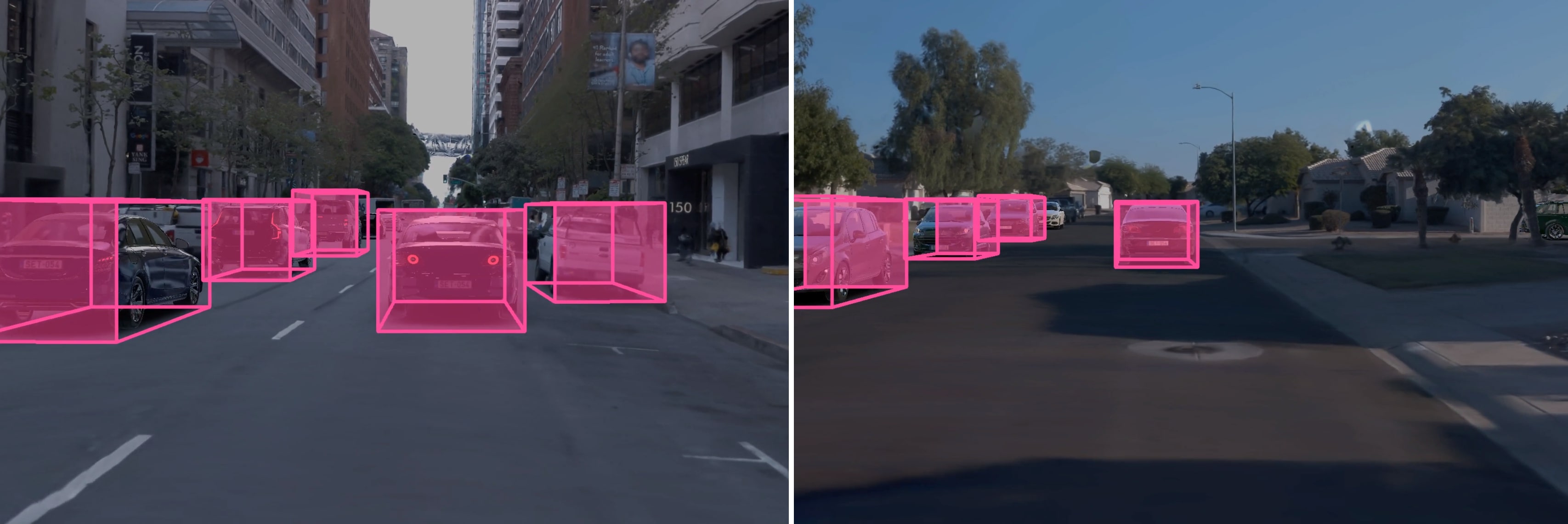}
    \caption{DEVIANT model detections on images rendered using our hybrid method.}
    \label{fig:detections_3dod_2}
\end{figure*}

\begin{figure*}
    \centering
    \includegraphics[width=1.0\linewidth]{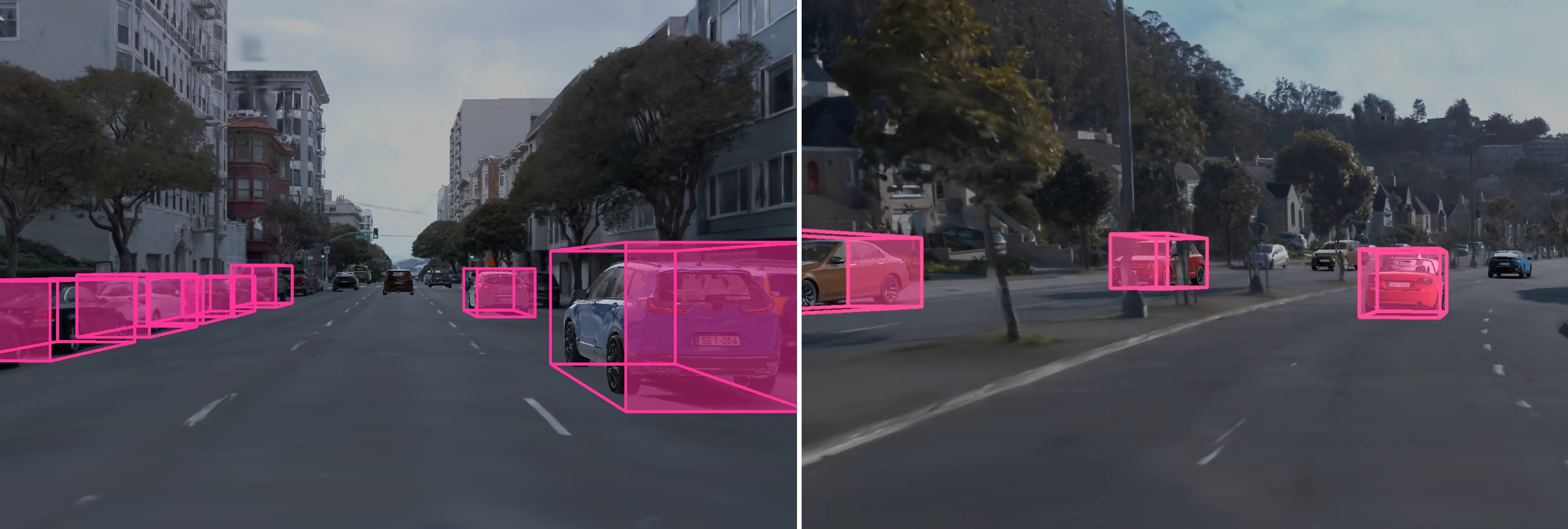}
    \caption{DEVIANT model detections on images rendered using our hybrid method.}
    \label{fig:detections_3dod_3}
\end{figure*}

Figure \ref{fig:detections_3dod_2} and \ref{fig:detections_3dod_3} show detections from DEVIANT \cite{kumar2022deviant}, a publicly available monocular model, on scenarios generated using our method. The reconstruction model rendered the static environment, while the mesh-based rendering engine introduced dynamic vehicles. As shown, the model successfully detects vehicles from both rendering methods, suggesting that no significant domain gap is introduced. Distant objects are not recognized due to model limitations.

We evaluated how the reconstruction quality of novel view synthesis impacts the model’s detection performance. Specifically, we compared detection results on the original trajectory with those from novel views where the ego-trajectory was shifted left by one, two, and three meters. This allows us to assess whether high-quality reconstructions of significantly altered viewpoints remain useful for autonomous driving tasks.

\begin{figure*}
    \centering
    \begin{subfigure}{1.0\linewidth}
        \centering
        \includegraphics[width=1.0\linewidth]{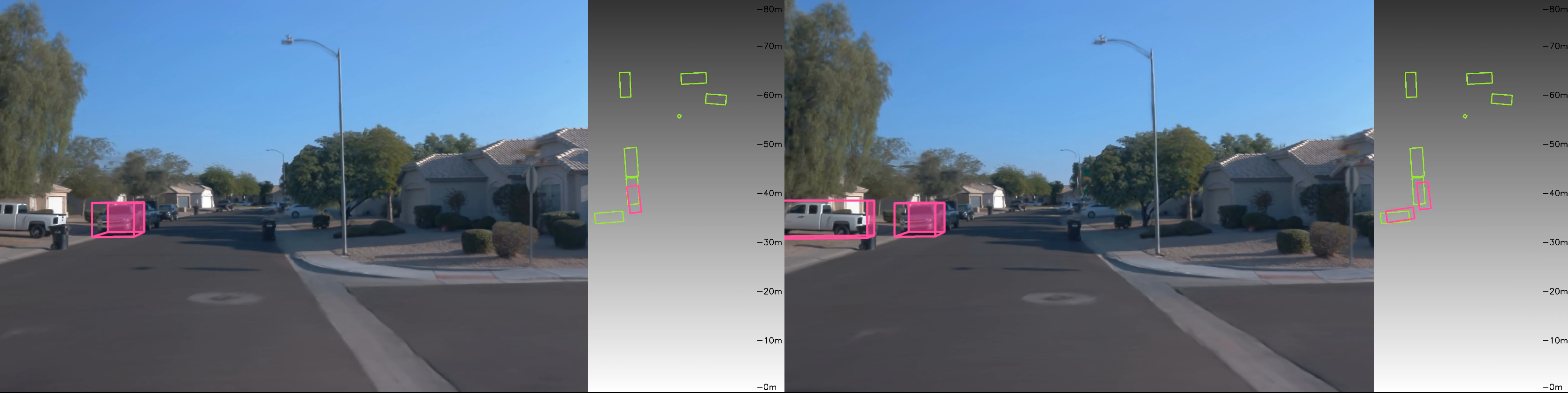}
        \caption{The ego-trajectory is unchanged (left) and is shifted to the left by one meter (right).}
        \label{fig:nvs2_01}
    \end{subfigure}
    
    \vspace{0.5cm} % Adds vertical space between the figures

    \begin{subfigure}{1.0\linewidth}
        \centering
        \includegraphics[width=1.0\linewidth]{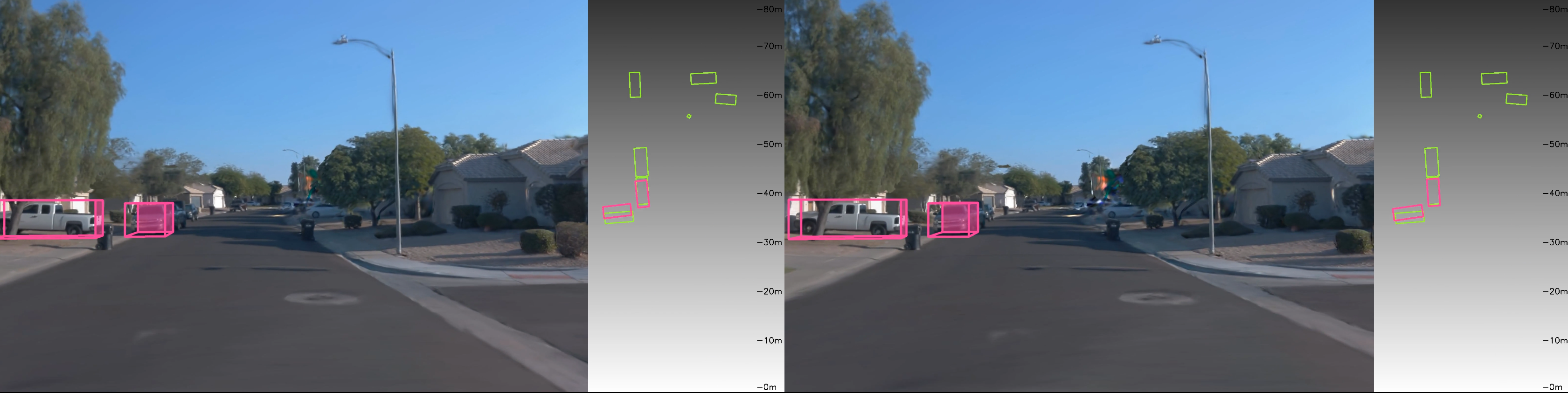}
        \caption{The ego-trajectory is shifted to the left by two meters (left) and three meters (right).}
        \label{fig:nvs2_23}
    \end{subfigure}
    
    \caption{Detections from DEVIANT model on a segment rendered from novel views.}
    \label{fig:nvs2_combined}
\end{figure*}

\begin{figure*}
    \centering
    \begin{subfigure}{1.0\linewidth}
        \centering
        \includegraphics[width=1.0\linewidth]{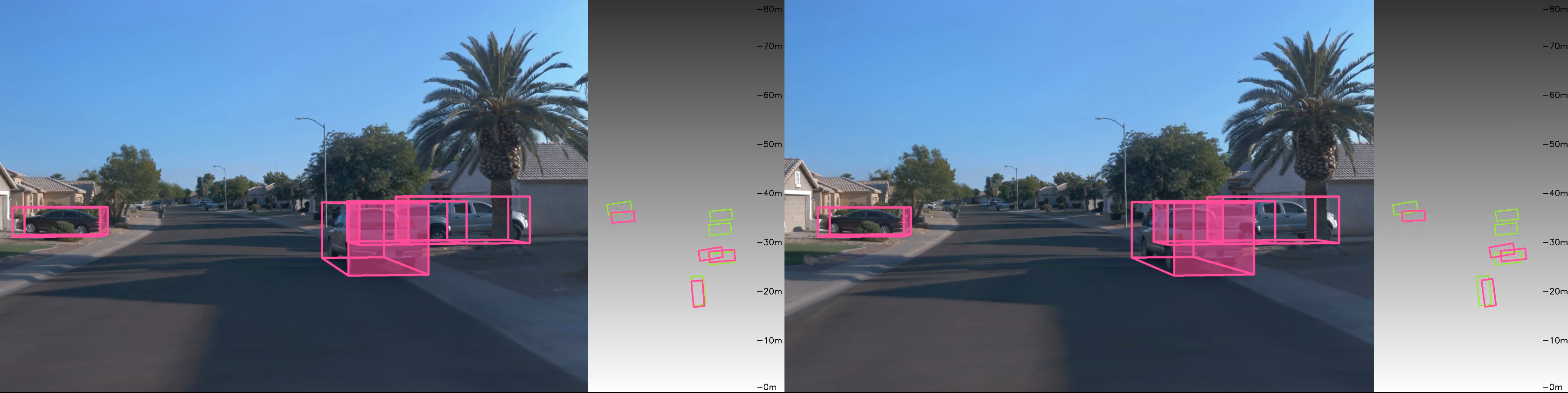}
        \caption{The ego-trajectory is unchanged (left) and is shifted to the left by one meter (right).}
        \label{fig:nvs1_01}
    \end{subfigure}
    
    \vspace{0.5cm} % Adds vertical space between the figures

    \begin{subfigure}{1.0\linewidth}
        \centering
        \includegraphics[width=1.0\linewidth]{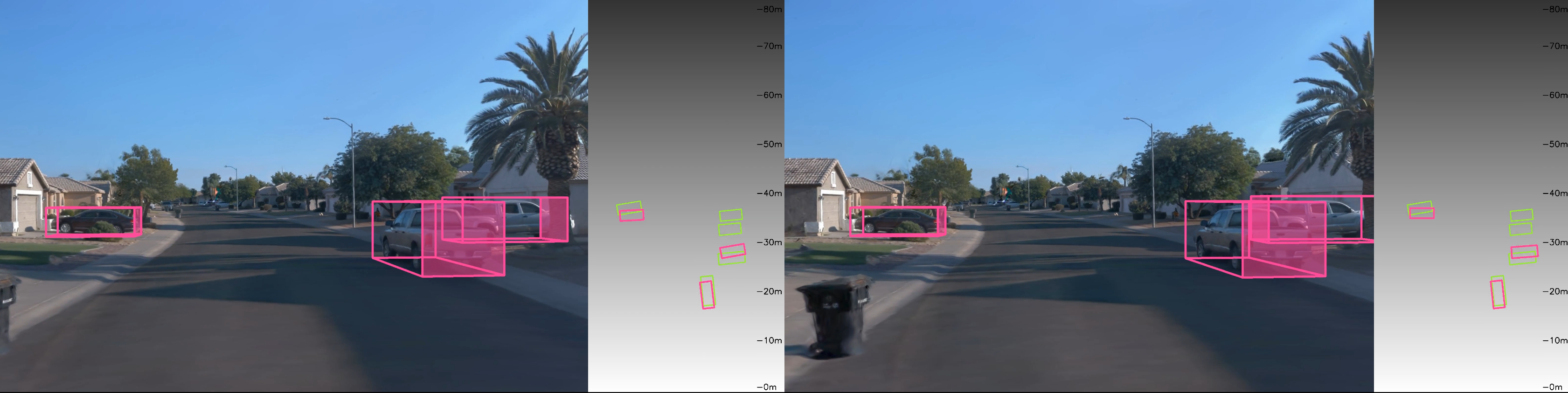}
        \caption{The ego-trajectory is shifted to the left by two meters (left) and three meters (right).}
        \label{fig:nvs1_23}
    \end{subfigure}
    
    \caption{Detections from DEVIANT model on a segment rendered from novel views.}
    \label{fig:nvs1_combined}
\end{figure*}

Figures \ref{fig:nvs1_combined} and \ref{fig:nvs2_combined} show 3D bounding boxes detected by the DEVIANT model on the same frame after these horizontal shifts. As observed, viewpoint changes can help reduce false positives by improving object visibility. However, they can also introduce false negatives when certain objects become less visible after the shift. These effects help explain the metric differences presented in the paper.

\section{System Performance and Limitations}
\subsection{Complexity}
%While the pipeline has multiple stages, each is modular, well-isolated, and produces interpretable outputs, enabling robust debugging and integration. Given valid input, the stages are stable and not error-prone. The tool is already used by industry partners and freely available for academic use, showing its practical viability beyond research.

While the pipeline involves multiple stages, each component is modular, well-isolated, and produces human-interpretable outputs. This modularity allows for effective debugging, maintenance, and integration, as each step can be validated independently, reducing the likelihood of error propagation. Assuming valid input data, the stages are robust and not inherently error-prone. Furthermore, the simulation tool is already in use by industrial partners as a closed-source solution and is available for academic use under a free license. This demonstrates the system’s practicality and usability beyond a purely research context.

\subsection{Block Boundary Artifacts}
\begin{figure}[h]
  \centering
   \includegraphics[width=1.0\linewidth]{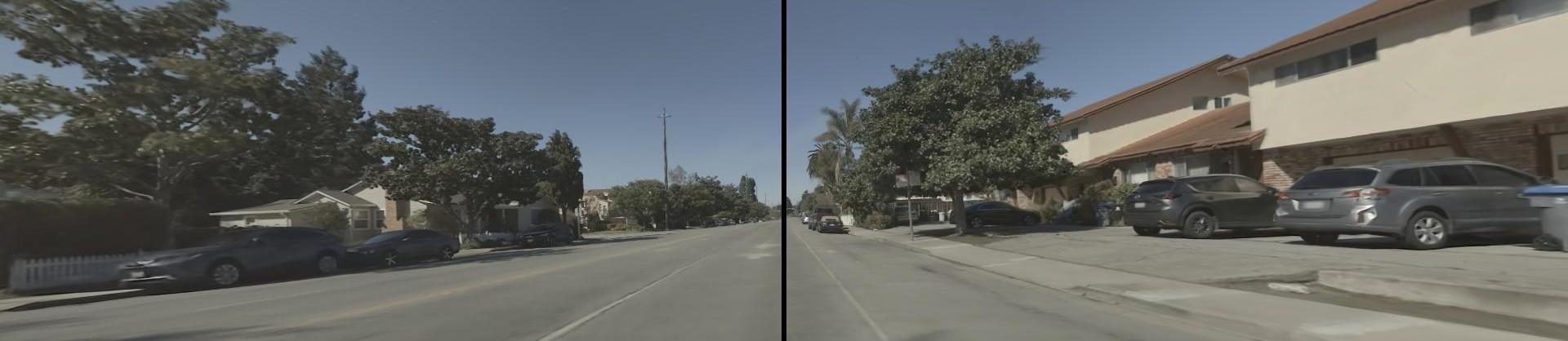}
   \caption{Images rendered from the same viewpoint from different blocks close to the block edge}
   \label{fig:block_edge}
\end{figure}

The difference in distant textures arose from a NeRF distance cutoff during initial point cloud generation. Removing this cutoff has significantly reduced block edge artifacts (seen in Fig. \ref{fig:block_edge}). To ensure a consistent sky, we propose training a single sky model from all images and freezing it during block training.

\subsection{Reconstruction efficiency/rendering speed}
Our method takes 12 ms to render a Full HD image, and a 64-beam LiDAR frame takes 17 ms on an NVIDIA RTX 3090. 
%\RED{ANSWER IS MISSING HERE!}
%%%%%%%%%

\end{document}